\documentclass[%
reprint,
superscriptaddress,
showpacs,preprintnumbers,
nofootinbib,
amsmath,amssymb,
aps,
pra,
showkeys,
floatfix,
twocolumn,
]{revtex4-1}
\usepackage{graphicx}
\usepackage[caption=false]{subfig}
\usepackage{dcolumn}
\usepackage{bm}
\usepackage{dsfont}
\usepackage{amsmath,amssymb,stackengine}
\usepackage{siunitx}
\usepackage[caption=false]{subfig}
\usepackage{braket}
\usepackage{listings}
\usepackage{xcolor}
\usepackage{hyperref}
\hypersetup{colorlinks=true,linkcolor=blue,urlcolor=blue,citecolor=blue}

\begin{document}
\title{Geometric phases in neutrino mixing}
\author{Manosh T. M.}
\email{tm.manosh@gmail.com}
\email{tm.manosh@cusat.ac.in}
\author{N. Shaji}
\affiliation{Department of Physics, Cochin University of Science and Technology, Kochi 682022, India.}
\author{Ramesh Babu Thayyullathil}
\author{Titus K. Mathew}
\affiliation{Department of Physics, Cochin University of Science and Technology, Kochi 682022, India.}
\affiliation{Centre for Particle Physics, Cochin University of Science and Technology, Kochi 682022, India.}
\date{\today}

\begin{abstract}
	Neutrinos can acquire both dynamic and geometric phases due to the non-trivial mixing between mass and flavour eigenstates. In this article, we derive the general expressions for all plausible gauge invariant diagonal and off-diagonal geometric phases in the three flavour neutrino model using the kinematic approach. We find that diagonal and higher order off-diagonal geometric phases are sensitive to the mass ordering and the Dirac CP violating phase $\delta$. We show that, third order off-diagonal geometric phase ($\Phi_{\mu e\tau}$) is invariant under any cyclic or non-cyclic permutations of flavour indices when the Dirac CP phase is zero. For non-zero $\delta$, we find that $\Phi_{\mu e\tau}(\delta)=\Phi_{e \mu \tau}(-\delta)$. Further, we explore the effects of matter background using a two flavour neutrino model and show that the diagonal geometric phase is either 0 or $\pi$ in the MSW resonance region and takes non-trivial values elsewhere. The transition between zero   and $\pi$ occurs at the point of complete oscillation inversion called the nodal point, where the diagonal geometric phase is not defined. Also, in two flavour approximations, two distinct diagonal geometric phases are co-functions with respect to the mixing angle. Finally, in the two flavour model, we show that the only second order off-diagonal geometric phase is a topological invariant quantity and is always $\pi$.
\end{abstract}
\keywords{Geometric phases, neutrinos, mass ordering, CP violation}
\maketitle

\section{\label{Sec.I}Introduction}

Many experiments \cite{PhysRevD.103.L011101,PhysRevLett.123.151803} support the theory of mixing between flavour and mass eigenstates as the underlying description for neutrino oscillations. The Pontecorvo-Maki-Nakagawa-Sakata (PMNS) formalism~\cite{osti_4349231,10.1143/PTP.28.870} 
describes the amplitude of flavour oscillation with three mixing angles ($\theta_{12},\,\theta_{13},\,\theta_{23}$) and CP violating phases. Furthermore, the frequencies of flavour oscillations are proportional to mass squared differences ($\Delta m^2_{21},\, \Delta m^2_{31}$)~\cite{RevModPhys.82.2701}. Two objectives of ongoing neutrino experiments are to resolve the mass ordering (hierarchy) problem \cite{PhysRevD.101.032006}, which is to settle the sign of $\Delta m^2_{31}$ and measure the CP violating phases~\cite{Abe2020} that appear in the PMNS formalism. There are several approaches to address these problems \cite{PhysRevD.101.093001,PhysRevD.72.013009,PhysRevD.86.113011,PhysRevD.95.053002,PhysRevD.104.015038}. This article explores geometric phases arising in neutrino oscillation to address the mass ordering problem and the Dirac CP violating phase ($\delta$). 

In I956, in a classical context, S. Pancharatnam \cite{Pancharatnam1956} showed the existence of a phase of geometric origin that depends on the path traced by the polarisation state of light on a Poincar\'{e} sphere. Around three decades later, M.~Berry~\cite{doi:10.1098/rspa.1984.0023} discovered a geometric phase factor accompanying quantum adiabatic changes in addition to the usual dynamic phase. S.~Ramaseshan and R. Nityananda \cite{10.2307/24090242} recognised a correspondence between the works of Pancharatnam and Berry. Later, Berry himself showed that the Pancharatnam phase is an optical analogue of the phase accompanying quantum adiabatic changes \cite{doi:10.1080/09500348714551321}. Y. Aharonov and J. Anandan \cite{PhysRevLett.58.1593} generalised geometric phases to non-adiabatic evolution, which then J. Samuel and R. Bhandari \cite{PhysRevLett.60.2339} presented in a more concrete framework, establishing the gauge invariance of geometric phases. Reviews by Cohen \textit{et al.} \cite{Cohen2019} and Jisha~\textit{et~al.}~\cite{https://doi.org/10.1002/lpor.202100003} present details about recent developments on the topic. We will revisit the details in Sec. \ref{Sec.II}.

N. Nakagawa \cite{NAKAGAWA1987145} was the first to consider geometric phase in neutrinos propagating through matter using higher order adiabatic approximations. Later, V.~Naumov~\cite{NAUMOV1994351} showed the existence of non-trivial geometric phases in the three flavour neutrino oscillation with a slowly varying matter environment. The work by X.~G.~He~\textit{et~al.}~\cite{PhysRevD.72.053012} demands a background with at least two independent matter densities and Dirac CP violating phases for neutrinos to generate a geometric phase. They also show that Majorana phases play no role in generating a geometric phase. Several authors followed the same arguments, which need varying external parameters to have non-trivial geometric phases \cite{JOSHI2016135,PhysRevD.95.043003,PhysRevD.96.096004}.

Alternatively, in 1993, R. Simon and N. Mukunda~\cite{PhysRevLett.70.880} introduced geometric phases as Bargmann invariants. This led to the general formalism known as the quantum kinematic approach to calculate geometric phases \cite{MUKUNDA1993205, PhysRevA.60.3397}. This approach is useful in exploring gauge invariant quantities arising in the kinematics, even in the absence of varying external parameters. Blasone~\textit{et~al.}~\cite{BLASONE1999262} chose this route and illustrated the gauge structure in neutrino oscillation as in the case of Aharonov-Bohm effect \cite{PhysRev.115.485} for a local gauge transformation. Wang~\textit{et~al.}~\cite{PhysRevD.63.053003} extended this approach to non-cyclic phases, which Dixit~\textit{et al.}~\cite{Dixit_2018} applied on the mass ordering problem.

Motivated from the kinematic approach, we explore gauge invariant geometric phases in the three flavour neutrino oscillation model. We derive the general formulae for all plausible gauge invariant geometric phase factors and discuss their sensitivity over mass ordering and Dirac CP phase. We explore the effects of matter potential using a two flavour neutrino model and illustrate the features of corresponding diagonal and off-diagonal geometric phases. We show that, in two flavour approximation, the only second order off-diagonal geometric phase is a topological quantity and is always $\pi$. While, diagonal geometric phase takes values 0 or $\pi$ only under bi-maximal mixing. Milestone article by P.~Mehta~\cite{PhysRevD.79.096013} was the first to discuss the difference between phases of topological and geometric origin in the context of neutrino oscillation.

We organise the paper as follows. In Sec. \ref{Sec.II} we discuss the details of diagonal and off-diagonal geometric phases. In Sec. \ref{Sec.III} we introduce the notations and conventions followed to discuss neutrino oscillation. In Sec.~\ref{Sec.IV} we derive general expressions for all geometric phases in neutrinos and investigate the effects of matter potential with a two flavour model. Finally, we summarise the results in Sec. \ref{Sec.V}.

\section{\label{Sec.II}Geometric phases}
 
Berry's discovery \cite{doi:10.1098/rspa.1984.0023} of geometric phase in quantum systems led to many debates, and discussions~\cite{PhysRevLett.89.220404, PhysRevLett.108.033601, PhysRevA.93.033823, Gasparinettie1501732}. Immediately, the notion of geometric phase got extended to mixed states \cite{UHLMANN1986229,PhysRevLett.85.2845}, open systems \cite{PhysRevLett.90.160402}, entangled systems \cite{PhysRevA.62.022109}, composite systems \cite{PhysRevLett.92.150406} and design of logic gates~\cite{Leek1889}. A breakthrough in the field came with the work of Simon and Mukunda \cite{MUKUNDA1993205}, where they gave a quantum kinematic description to geometric phases. Tong \textit{et al.} \cite{PhysRevLett.93.080405} extended the kinematic approach to non-unitary evolution with mixed states. In the kinematic method, one can rewrite geometric phases as Bargmann invariant and relax the assumption of adiabatic limit even for off-diagonal geometric phase \cite{PhysRevA.65.012102}. Recent work by Khan~\textit{et~al.}~\cite{KHAN2021100061} on geometric phases outlines a new perspective on interference in phase space. Therefore, studies on geometric phases can give new insights into old problems. In this section, we discuss the details of diagonal and off-diagonal geometric phases.

\subsection{Diagonal geometric phases ($\Phi_i$)}
Consider an $N$-dimensional quantum system and let $\{\ket{\phi_n}\}$ be a set of complete orthonormal basis.  Let $\ket{\phi_i}$ be an initial state, then the state after a time $t$ is,
\begin{equation}
\ket{\psi_i(t)}=\hat{\mathds{U}}(t)\ket{\phi_i}=\sum_{n=1}^{N}c_n(t)\ket{\phi_n}.
\label{generalstate}
\end{equation}
Here, $\hat{\mathds{U}}(t)$ is the unitary time evolution operator for a given Hamiltonian $\hat{H}$. To recognize the final state produced from a specific initial state $\ket{\phi_i}$ we put the subscript to the final state as $\ket{\psi_i(t)}$. 

In the kinematic approach, total phase is the sum of dynamic and geometric phases. One must then eliminate the dynamic phase from the total phase to recover geometric phase factors. To off-set the dynamic phase factor we use the definition of parallel transported state given by \cite{PhysRevLett.60.2339},
\begin{equation}
\ket{\psi_i^{\parallel}(t)}=\exp\left[-\int_{0}^{t}\bra{\psi_i(t')}\partial_{t'}\ket{\psi_i(t')}dt'\right]\ket{\psi_i(t)}.
\label{parallel}
\end{equation}
Here, the exponential factor removes the dynamic phase and by the definition of parallel transport, we have, $\ket{\psi_i^{\parallel}(0)}=\ket{\psi_i(0)}$. Since the parallel transported states are devoid of all dynamic factors, inner product between them will generate quantities with geometric phase factors. Now, we define our inner product as,
\begin{equation}
\sigma_{ii}=\braket{\psi_i^{\parallel}(0)|\psi_i^{\parallel}(t)}.
\label{sigmaii}
\end{equation}
We may call $\sigma_{ii}$ the first order geometric phase factor, as there is only one index. On considering a local gauge transformation of the form:
\begin{equation}
\ket{\psi_i(t)}\rightarrow\exp\left[i\zeta_i(t)\right]\ket{\psi_i(t)},
\label{eq:localgauge}
\end{equation}
we can easily verify that $\sigma_{ii}$ is gauge invariant and hence a measurable physical quantity. We now define the diagonal geometric phase as,
\begin{equation}
\Phi_i=\arg\left[\sigma_{ii}\right].
\end{equation} 
The word ``diagonal" corresponds to the identical indices of $\sigma_{ii}$ similar to diagonal elements in a matrix. 

\subsection{Off-diagonal geometric phases ($\Phi_{ij}$)} 
The expression of final state given by Eq.~(\ref{generalstate}) can have components orthonormal to the initial state. When we take the inner product between the initial and the final states we lose information about all other orthonormal components. Things worsen if the final state itself is orthonormal to the initial state. Then the inner product becomes zero, and the phase is no longer defined. To resolve this problem, N. Manini and F. Pistolesi \cite{PhysRevLett.85.3067} introduced the idea of off-diagonal geometric phases for adiabatic evolution. Further, Mukunda \textit{et al.} \cite{PhysRevA.65.012102} generalised these results by relaxing the adiabatic constraints. Thus we need gauge invariant off-diagonal and diagonal geometric phase factors to give the complete description of geometric phases. Later, Filipp \textit{et al.} \cite{PhysRevLett.90.050403} extended the analysis to mixed states, and Wong \textit{et al.} \cite{PhysRevLett.94.070406} introduced the notion of projective measurement to compute geometric phase between any two states.

Let $\ket{\phi_i}$ and $\ket{\phi_j}$ be two orthonormal states and let $\ket{\phi_j}$ be the initial state. Then using the notion of parallel transport defined earlier, we can construct an inner product of the form,
\begin{equation}
\sigma_{ij}=\braket{\psi_i^{\parallel}(0)|\psi_j^{\parallel}(t)}.
\label{sigmaab}
\end{equation}
On considering a local gauge transformation given by Eq.~(\ref{eq:localgauge}), we get
\begin{equation}
\sigma_{ij}\rightarrow\sigma_{ij}\exp\left\{i\left[\zeta_i(t)-\zeta_j(t)\right]\right\}.
\end{equation}
Thus, $\sigma_{ij}$ is not gauge invariant and hence immeasurable. Similarly, 
\begin{equation}
\sigma_{ji}\rightarrow\sigma_{ji}\exp\left\{i\left[\zeta_j(t)-\zeta_i(t)\right]\right\},
\end{equation}
is also immeasurable. Fortunately, since $\sigma_{ij}$ and $\sigma_{ji}$ are complex scalar quantities, we could multiply them and define $\sigma_{ij}\sigma_{ji}$, which is gauge invariant under Eq. (\ref{eq:localgauge}) and hence represents a physical measurable quantity. With this, we now define the off-diagonal geometric phase as,
\begin{equation}
\Phi_{ij}=\arg\left[\sigma_{ij}\sigma_{ji}\right].
\end{equation}
Here, $\Phi_{ij}$ is gauge invariant and measurable. Since there are two distinct indices, we call it the second order off-diagonal geometric phase. The permutation of two indices are immaterial as, $\sigma_{ij}\sigma_{ji}=\sigma_{ji}\sigma_{ij}$, hence, $\Phi_{ij}=\Phi_{ji}$.  One can then generalise this idea and calculate higher order off-diagonal geometric phases as,
\begin{equation}
\Phi_{ijk...pq}=\arg\left[\sigma_{ij}\sigma_{jk}...\sigma_{pq}\sigma_{qi}\right],
\end{equation}
where no indices repeat. If there are repeating indices, we could decompose them in terms of lower order off-diagonal and diagonal geometric phase factors. 

Although the cyclic permutations of indices are immaterial, non-cyclic permutations give additional geometric phases when there is CP violation. Then naturally, third and higher order off-diagonal geometric phase factors carry such non-trivial dependencies resulting in additional geometric phase factors, which is one of the central results of this article. We shall show this explicitly by considering the three flavour neutrino model. 

\section{\label{Sec.III}Neutrino Hamiltonian}

Neutrinos come with a flavour tag based on the sister lepton ($e,\mu,\tau$) produced along with it. Interestingly, a neutrino produced with a specific flavour oscillates to a different flavour as it evolves. This phenomenon based on the mixing between flavour and mass eigenstates via the Pontecorvo-Maki-Nakagawa-Sakata (PMNS) formalism \cite{osti_4349231,10.1143/PTP.28.870} is well established. The motivation of CP violating phase in the PMNS formalism originally comes from the CP violation in the renormalizable theory of weak interaction due to M. Kobayashi and T. Maskawa~\cite{10.1143/PTP.49.652}. Additionally, the medium through which neutrinos propagates can also influence flavour transitions, which Wolfenstein~\cite{PhysRevD.17.2369} explained. Later, Mikheyev and Smirnov \cite{Mikheyev1986} discovered the resonance effect in matter induced flavour transition, known as the Mikheyev-Smirnov-Wolfenstein (MSW) effect. In the following subsections, we construct the neutrino flavour Hamiltonian for three and two flavour models.\\

\subsection{Three flavour model}
In the standard three flavour neutrino model we have three active neutrinos: the electron neutrino $\ket{\nu_{e}}$, the muon neutrino $\ket{\nu_{\mu}}$ and the tau neutrino $\ket{\nu_{\tau}}$. This model is a three level quantum system for which we can take a complete orthonormal basis in which the basis vectors represents the flavour states as,
\begin{equation}
\ket{\nu_{e}}=\begin{pmatrix}
1\\0\\0
\end{pmatrix}\!\!,\,
\ket{\nu_{\mu}}=\begin{pmatrix}
0\\1\\0
\end{pmatrix} \text{and} 
\ket{\nu_{\tau}}=\begin{pmatrix}
0\\0\\1
\end{pmatrix}.
\end{equation}
The Hamiltonian (with $\hbar=c=1$) in flavour basis is,
\begin{equation}
\hat{H}_F=U\begin{pmatrix}
0&0&0\\
0&\frac{\Delta m^2_{21}}{2E}&0\\
0&0&\frac{\Delta m^2_{31}}{2E}
\end{pmatrix}U^{\dagger}+\begin{pmatrix}
V_{CC}&0&0\\
0&0&0\\
0&0&0
\end{pmatrix}.
\label{3flavourHF}
\end{equation}
$U$ is the PMNS matrix given as,
\begin{widetext}
	\begin{align}U&={\begin{pmatrix}U_{e1}&U_{e2}&U_{e3}\\U_{\mu 1}&U_{\mu 2}&U_{\mu 3}\\U_{\tau 1}&U_{\tau 2}&U_{\tau 3}\end{pmatrix}}={\begin{pmatrix}c_{12}c_{13}&s_{12}c_{13}&s_{13}e^{-i\delta }\\-s_{12}c_{23}-c_{12}s_{23}s_{13}e^{i\delta }&c_{12}c_{23}-s_{12}s_{23}s_{13}e^{i\delta }&s_{23}c_{13}\\s_{12}s_{23}-c_{12}c_{23}s_{13}e^{i\delta }&-c_{12}s_{23}-s_{12}c_{23}s_{13}e^{i\delta }&c_{23}c_{13}\end{pmatrix}}.
	\label{PMNS}
	\end{align}
\end{widetext}
Here, $s_{ij}=\sin\theta_{ij}$ and $c_{ij}=\cos\theta_{ij}$, where, $\theta_{12}$, $\theta_{13}$, $\theta_{23}$ are the three mixing angles and $\delta$ is the Dirac CP violating phase. Further, $\Delta m^2_{ij}=m^2_i-m^2_j$ is the difference between the square of masses of $i^{\text{th}}$ and $j^{\text{th}}$ mass eigenstate, $E$ is the total energy of neutrino and $V_{CC}=\sqrt{2}G_FN_e$ is the charged current interaction potential due to the medium through which neutrinos propagate. Here $G_F$ is the Fermi constant and $N_e$ is electron number density. If we consider non-standard interactions (NSI), the interaction energy matrix in Eq. (\ref{3flavourHF}) can have more non-zero entries. In this analysis, we will not consider any NSI or additional CP violating Majorana phases. Several authors~\cite{CAPOLUPO2018216,LU2021136376, Capolupo:2021enm, Johns:2021ets} discuss the significance of Majorana phases in the context of geometric phase.

Finally, given an initial flavour state $\ket{\nu_{\alpha}}$ the state at a later time is,
\begin{equation}
\ket{\nu_{\alpha}(t)}=\exp\left(-i\int_{0}^{t}\hat{H}_Fdt'\right)\ket{\nu_{\alpha}}.
\label{nuoft}
\end{equation}
The subscript $\alpha$ in the final state is to recognise the initial flavour from which it came. One must not confuse $\ket{\nu_{\alpha}(t)}$ as $c_{\alpha}(t)\ket{\nu_{\alpha}}$; the first represents the complete state at time $t$ while the second corresponds to a flavour component of the state at time $t$. 

\subsection{Two flavour model}
In most practical situations we can approximate the three flavour neutrino model to two flavour neutrino model. This approximation makes it convenient to analyse the effects of matter potential. Under this approximation, Eq.  (\ref{3flavourHF}) reduces to,
\begin{equation}
\hat{H}_F=U\begin{pmatrix}
0&0\\
0&\frac{\Delta m^2}{2E}
\end{pmatrix}U^{\dagger}+\begin{pmatrix}
V_{CC}&0\\
0&0
\end{pmatrix}.
\label{2flavourHF}
\end{equation}
and the PMNS matrix becomes,
\begin{equation}
U=\begin{pmatrix}
\cos\theta&\sin\theta\\
-\sin\theta&\cos\theta
\end{pmatrix}.
\end{equation}
Here, $\Delta m^2$ can be either $\Delta m^2_{21}$ or $\Delta m^2_{31}$ and $\theta$ can be either $\theta_{12}$ or $\theta_{23}$, depending on the oscillation channel. In the two flavour approximation we have two main oscillation channels. One is $\ket{\nu_{\mu}}\rightarrow\ket{\nu_{\tau}}$ for small $L/E$ values and the other is $\ket{\nu_{e}}\rightarrow\ket{\nu_{\mu/\tau}}$ for large values of $L/E$. Where, $L$ is the distance travelled by the neutrino.

 If we assume, $\ket{\nu_{\mu}}\rightarrow\ket{\nu_{\tau}}$ as the two flavour approximation, the matter potential due to charged current interaction becomes zero. Else, we need a muon or tau rich medium which is not the usual case. Such situations could exist in the early phase of the Universe. Instead, if we consider $\ket{\nu_{e}}\rightarrow\ket{\nu_{\mu/\tau}}$, the corresponding matter potential will be non-zero in electron rich medium such as Earth and Sun. By including the matter potential one can redefine $\theta$ and $\Delta m^2$ to $\theta_M$ and $\Delta m^2_M$ as,
\begin{equation}
\Delta m^2_M=\Delta m^2\sqrt{\left(\cos(2\theta)-\frac{2EV_{CC}}{\Delta m^2}\right)^2+\sin^2(2\theta)},
\label{MSWmass}
\end{equation}
and
\begin{equation}
\tan(2\theta_M)=\tan(2\theta)\left[{1-\frac{2EV_{CC}}{\Delta m^2\cos(2\theta)}}\right]^{-1},
\label{MSWangle}
\end{equation}
such that, all the equations derived using the Hamiltonian retains the same form. Here we follow the convention in which,  $V_{CC}$ is positive for neutrinos and negative for anti-neutrinos.  In Eqs. (\ref{MSWmass}) and (\ref{MSWangle}), the factor $\left[1-2EV_{CC}/(\Delta m^2\cos(2\theta))\right]$ makes  $\Delta m^2_M$ and $\theta_M$ sensitive to the sign of $\Delta m^2$. For $2EV_{CC}=\Delta m^2\cos(2\theta)$, the mixing angle $\theta_M$ becomes $\pi/4$ resulting the resonance known as Mikheyev-Smirnov-Wolfenstein (MSW) resonance \cite{PhysRevD.17.2369,Mikheyev1986}. Hence, we will use the two flavour approximation to address the effects of matter potential. 

\subsection{Units and conventions \label{UC}}

It is convenient to express $E$ in GeV and $\Delta m^2_{ij}$ in eV$^2$ to have a quick reference with most neutrino experiments.  Since neutrinos are mostly relativistic and extremely low mass, we approximate $ct\approx L$, where $L$ is the distance travelled in kilometres, and $c$ is the speed of light in vacuum.  Since $ct\approx L$, we have $cdt\approx dL$. Hence, we will replace all the time integrals with distance integrals, and the limits become $0$ and $L$. In these units, we have,
\begin{equation}
\frac{\Delta m^2_{ij}c^3L}{4\hbar E}\approx1.27\times\frac{\Delta m^2_{ij}L}{E}.
\end{equation}
In our expressions we shall have,
\begin{equation}
\frac{\Delta m^2_{ij}c^3L}{2\hbar E}\approx2.54\times\frac{\Delta m^2_{ij}L}{E}:=\frac{\Delta M^2_{ij}L}{E}.
\end{equation}
We define $\Delta M^2_{ij}:=2.54\times\Delta m^2_{ij}$ to simplify the expressions.

 \section{\label{Sec.IV}Geometric phases in neutrinos}
In this section we derive the exact general formulae for all plausible geometric phases in three and two flavour neutrino model and investigate the effects due to matter potential using the two flavour model. 

\subsection{Diagonal geometric phases in neutrinos}

In three flavour neutrino model we can define three independent diagonal geometric phases namely, $\Phi_{e}$, $\Phi_{\mu}$ and $\Phi_{\tau}$. In general, we can take the flavour as $\alpha\in\{e,\mu,\tau\}$ and compute $\Phi_{\alpha}$. For simplicity, we assume $V_{CC}=0$. From Eqs. (\ref{parallel}) and (\ref{sigmaii}), we have,
\begin{equation}
\sigma_{\alpha\alpha}=\braket{\nu_{\alpha}^{\parallel}(0)|\nu_{\alpha}^{\parallel}(L)}.
\label{sigmaaa}
\end{equation}
Here, $\ket{\nu_{\alpha}(0)}=\ket{\nu_{\alpha}}$ and $\ket{\nu_{\alpha}(L)}$ is the total state after propagating a distance $L$. Then, using the definition of parallel transport in Eq. (\ref{parallel}), we can rewrite Eq. (\ref{sigmaaa}) as,
\begin{align}
&\sigma_{\alpha\alpha}=\bra{\nu_{\alpha}}\exp\left[-\int_{0}^{L}dL'\bra{\nu_{\alpha}(L')}\partial_{L'}\ket{\nu_{\alpha}(L')}\right]\ket{\nu_{\alpha}(L)}\nonumber\\
&=\braket{\nu_{\alpha}|\nu_{\alpha}(L)}\exp\left[-\int_{0}^{L}dL'\bra{\nu_{\alpha}(L')}\partial_{L'}\ket{\nu_{\alpha}(L')}\right].
\label{newsigmaaa}
\end{align}
Since our flavour Hamiltonian is time independent and we took the approximation $ct\approx L$, the state after travelling a distance $L$ is,
\begin{align}
\ket{\nu_{\alpha}(L)}=\exp\left(-i\hat{H}_FL\right)\ket{\nu_{\alpha}}.
\end{align}
Now, the inner product in Eq. (\ref{newsigmaaa}) takes the form,
\begin{align}
\braket{\nu_{\alpha}|\nu_{\alpha}(L)}\nonumber = ~&|U_{\alpha1}|^2  +  |U_{\alpha2}|^2 \exp\left\{{- \frac{i \Delta{M^2_{21}} L}{E}}\right\}  \\&+  |U_{\alpha3}|^2 \exp\left\{{- \frac{i \Delta{M^2_{31}} L}{E}}\right\},
\end{align}
where, $|U_{\alpha i}|^2=U_{\alpha i}U_{\alpha i}^*$. $U_{\alpha i}$ corresponds to the matrix elements of the PMNS matrix given in Eq. (\ref{PMNS}) and $U_{\alpha i}^*$ is its complex conjugate.

Now we can evaluate the integral in the exponential part of Eq. (\ref{newsigmaaa}). Since $\left[\hat{H}_F,\exp\left(-i\hat{H}_FL\right)\right]=0$, we have,
\begin{align}
-\int_{0}^{L}dL'&\braket{\nu_{\alpha}(L')|\partial_{L'}|\nu_{\alpha}(L')}=i\bra{\nu_{\alpha}(0)}\hat{H}_F\ket{\nu_{\alpha}(0)}L\nonumber\\
&=i\frac{L}{E} \left( |U_{\alpha 2}|^2 \Delta{M^2_{21}} + |U_{\alpha 3}|^2 \Delta{M^2_{31}} \right).
\label{dynamicaloffset}
\end{align}
Then, 
\begin{align}
\sigma_{\alpha\alpha}=&\left(|U_{\alpha1}|^2  +  |U_{\alpha2}|^2 \exp\left\{{- \frac{i \Delta{M^2_{21}} L}{E}}\right\} \right.\nonumber \\&\left.+  |U_{\alpha3}|^2 \exp\left\{{- \frac{i \Delta{M^2_{31}} L}{E}}\right\}\right)\nonumber\\
&\times\exp\left[i\frac{L}{E} \left( |U_{\alpha 2}|^2 \Delta{M^2_{21}} + |U_{\alpha 3}|^2 \Delta{M^2_{31}} \right)\right]
\end{align}
Now the expression for diagonal geometric phase is,
\begin{equation}
\Phi_{\alpha}=\arg\left[\sigma_{\alpha\alpha}\right]=\arctan\left[\frac{\Im\left(\sigma_{\alpha\alpha}\right)}{\Re\left(\sigma_{\alpha\alpha}\right)}\right].
\label{dgp}
\end{equation}
where,
\begin{widetext}
	\begin{align}
	\Im\left(\sigma_{\alpha\alpha}\right)=&\left\{\sin{\left[\frac{L}{E}\left(  \Delta{M^2_{31}}(|U_{\alpha3}|^{2} -1)+  \Delta{M^2_{21}} \left|{U_{\alpha2}}\right|^{2} \right) \right]}|U_{\alpha3}|^{2}\right. + \sin{\left[\frac{L}{E} \left(  \Delta{M^2_{31}}|U_{\alpha3}|^{2} +  \Delta{M^2_{21}} (\left|{U_{\alpha2}}\right|^{2}-1) \right) \right]} \left|{U_{\alpha2}}\right|^{2} \nonumber\\&\left.+ \sin{\left[\frac{ L}{E} \left( \Delta{M^2_{31}}|U_{\alpha3}|^{2} + \Delta{M^2_{21}} \left|{U_{\alpha2}}\right|^{2}\right) \right]} \left|{U_{\alpha1}}\right|^{2}\right\}\bigg/\left|\sigma_{\alpha\alpha}\right|,\label{imdgp}\\
	\Re\left(\sigma_{\alpha\alpha}\right)=&\left\{\cos{\left[\frac{L}{E}\left(  \Delta{M^2_{31}}(|U_{\alpha3}|^{2} -1)+  \Delta{M^2_{21}} \left|{U_{\alpha2}}\right|^{2} \right) \right]}|U_{\alpha3}|^{2}\right. + \cos{\left[\frac{L}{E} \left(  \Delta{M^2_{31}}|U_{\alpha3}|^{2} +  \Delta{M^2_{21}} (\left|{U_{\alpha2}}\right|^{2}-1) \right) \right]} \left|{U_{\alpha2}}\right|^{2} \nonumber\\&\left.+ \cos{\left[\frac{ L}{E} \left( \Delta{M^2_{31}}|U_{\alpha3}|^{2} + \Delta{M^2_{21}} \left|{U_{\alpha2}}\right|^{2}\right) \right]} \left|{U_{\alpha1}}\right|^{2}\right\}\bigg/\left|\sigma_{\alpha\alpha}\right|.\label{redgp}
	\end{align}
\end{widetext}
Using Eq. (\ref{dgp}) to (\ref{redgp}), we can compute the diagonal geometric phases for each flavours $\alpha\in\{e,\mu,\tau\}$. 

In the above expressions, we have $\Delta M^2_{31}$, which is $2.54\times\Delta m^2_{31}$, appearing inside the arguments of sine and cosine functions. Depending on the sign of $\Delta m^2_{31}$, the factors $\Delta{M^2_{31}}(|U_{\alpha3}|^{2} -1)+  \Delta{M^2_{21}} \left|{U_{\alpha2}}\right|^{2}$, $\Delta{M^2_{31}}|U_{\alpha3}|^{2} +  \Delta{M^2_{21}} (\left|{U_{\alpha2}}\right|^{2}-1)$ and $\Delta{M^2_{31}}|U_{\alpha3}|^{2} + \Delta{M^2_{21}} \left|{U_{\alpha2}}\right|^{2}$ takes different values. Thus the sign of $\Delta M^2_{31}$ makes a notable difference in the value of $\sigma_{\alpha\alpha}$. Hence, it is clear that the diagonal geometric phases are sensitive to the mass ordering.

Furthermore, $\Phi_{\mu}$ and $\Phi_{\tau}$ are sensitive to the magnitude of the Dirac CP violating phases $\delta$, while $\Phi_{e}$ is not. For $\alpha=e$, the expression only has $|U_{e i}|^2$, and from the PMNS matrix in Eq. (\ref{PMNS}), all $|U_{e i}|^2$ are devoid of the phase $\delta$. Similarly, when $\alpha=\mu$ or $\tau$, from the PMNS matrix we have,

\begin{align}
|U_{\mu 1}|^2=&c_{12}^{2} s_{13}^{2} s_{23}^{2} + 2 c_{12} c_{23} s_{12} s_{13} s_{23} \cos{\delta} + c_{23}^{2} s_{12}^{2},\nonumber\\
|U_{\mu 2}|^2=&c_{12}^{2} c_{23}^{2} - 2 c_{12} c_{23} s_{12} s_{13} s_{23} \cos{\delta} + s_{12}^{2} s_{13}^{2} s_{23}^{2},\nonumber\\
|U_{\tau 1}|^2=&c_{12}^{2} c_{23}^{2} s_{13}^{2} - 2 c_{12} c_{23} s_{12} s_{13} s_{23} \cos{\delta} + s_{12}^{2} s_{23}^{2},\nonumber\\
|U_{\tau 2}|^2=& c_{12}^{2} s_{23}^{2} + 2 c_{12} c_{23} s_{12} s_{13} s_{23} \cos{ \delta } + c_{23}^{2} s_{12}^{2} s_{13}^{2}.
\end{align}
and $|U_{\alpha3}|^2$ has no dependency on $\delta$. Since $\cos\delta$ is an even function, the above expressions are not sensitive to the sign of $\delta$. Hence, the diagonal geometric phases $\Phi_{\mu}$ and $\Phi_{\tau}$ are only sensitive to the magnitude of $\delta$.

Further, we plot the diagonal geometric phase as a function of energy for a fixed distance of 812 km. The motivation for the choice of energy range and the distance comes from the NOvA experiment \cite{PhysRevLett.123.151803}. Further, we use the mean values reported by the Particle Data Group \cite{10.1093/ptep/ptaa104} as, $\theta_{12}$=$33.64^{\circ}$, $\theta_{13}$=$8.53^{\circ}$, $\theta_{23}$=$47.63^{\circ}$, $\Delta m^2_{21}$=$7.53\times10^{-5}$ eV$^2$, $|\Delta m^2_{32}|$=$2.54\times10^{-3}$ eV$^2$ and $\delta$=$1.36\pi$.

\begin{figure}[b]
\includegraphics[width=0.425\textwidth]{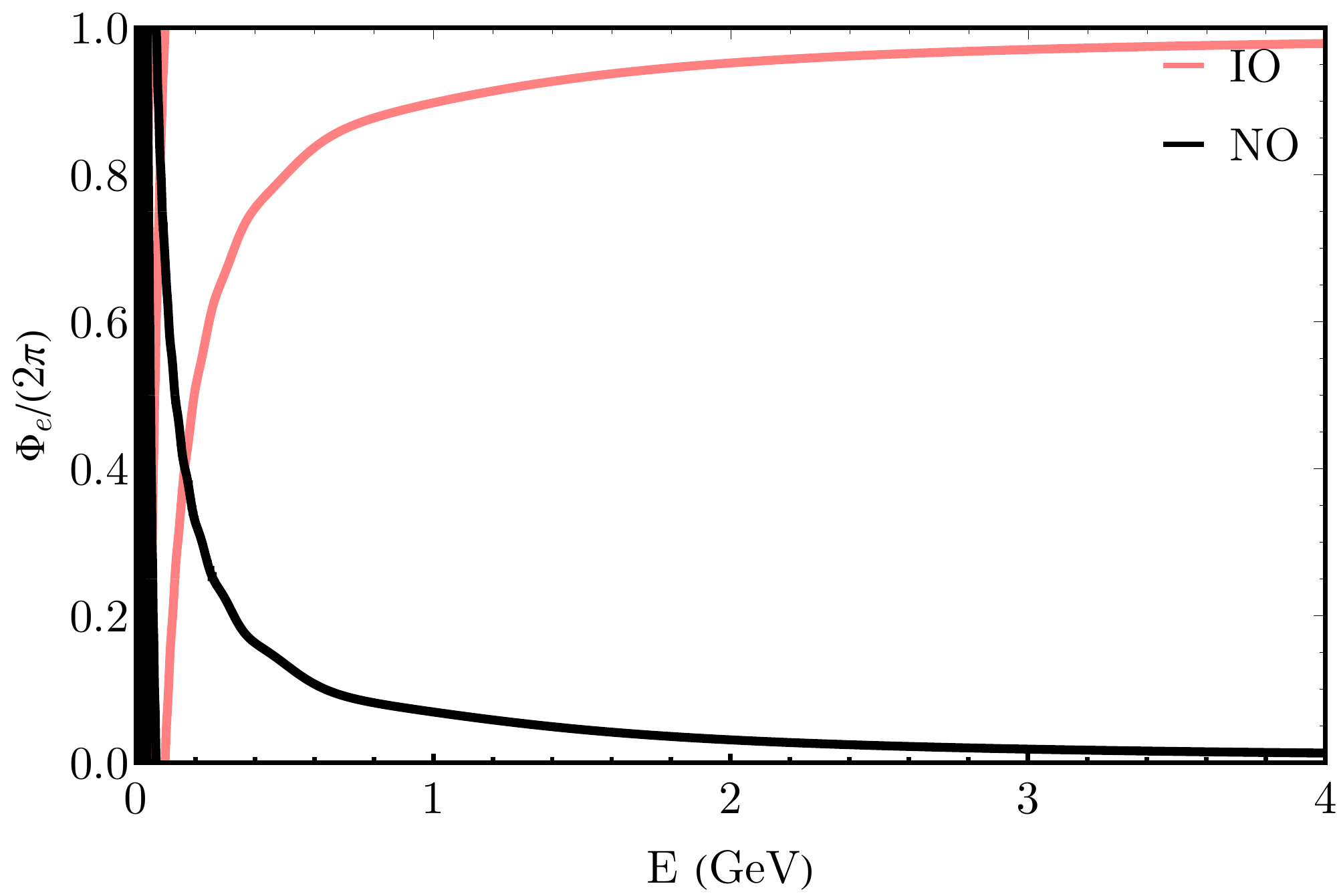}\\
\includegraphics[width=0.425\textwidth]{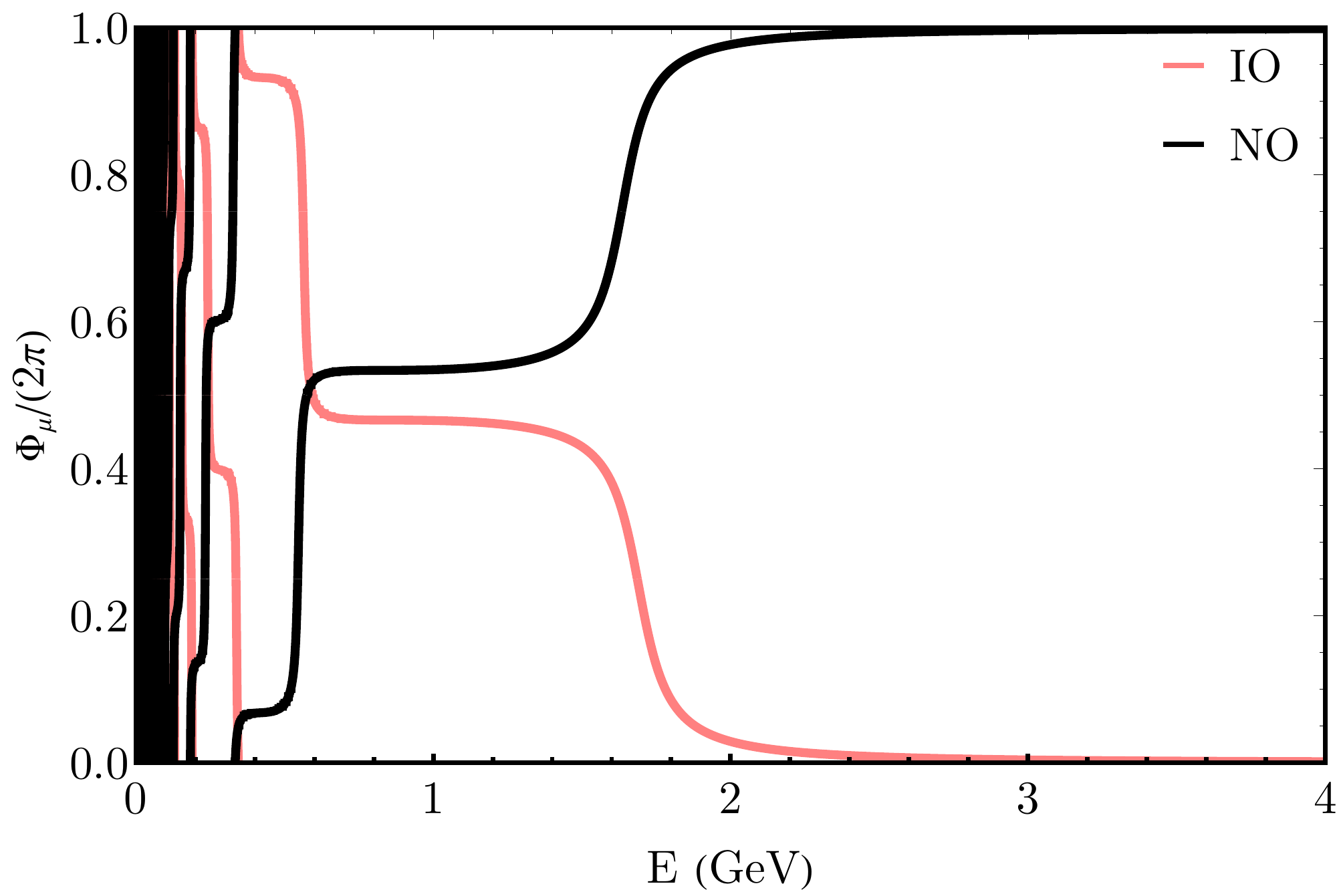}\\
\includegraphics[width=0.425\textwidth]{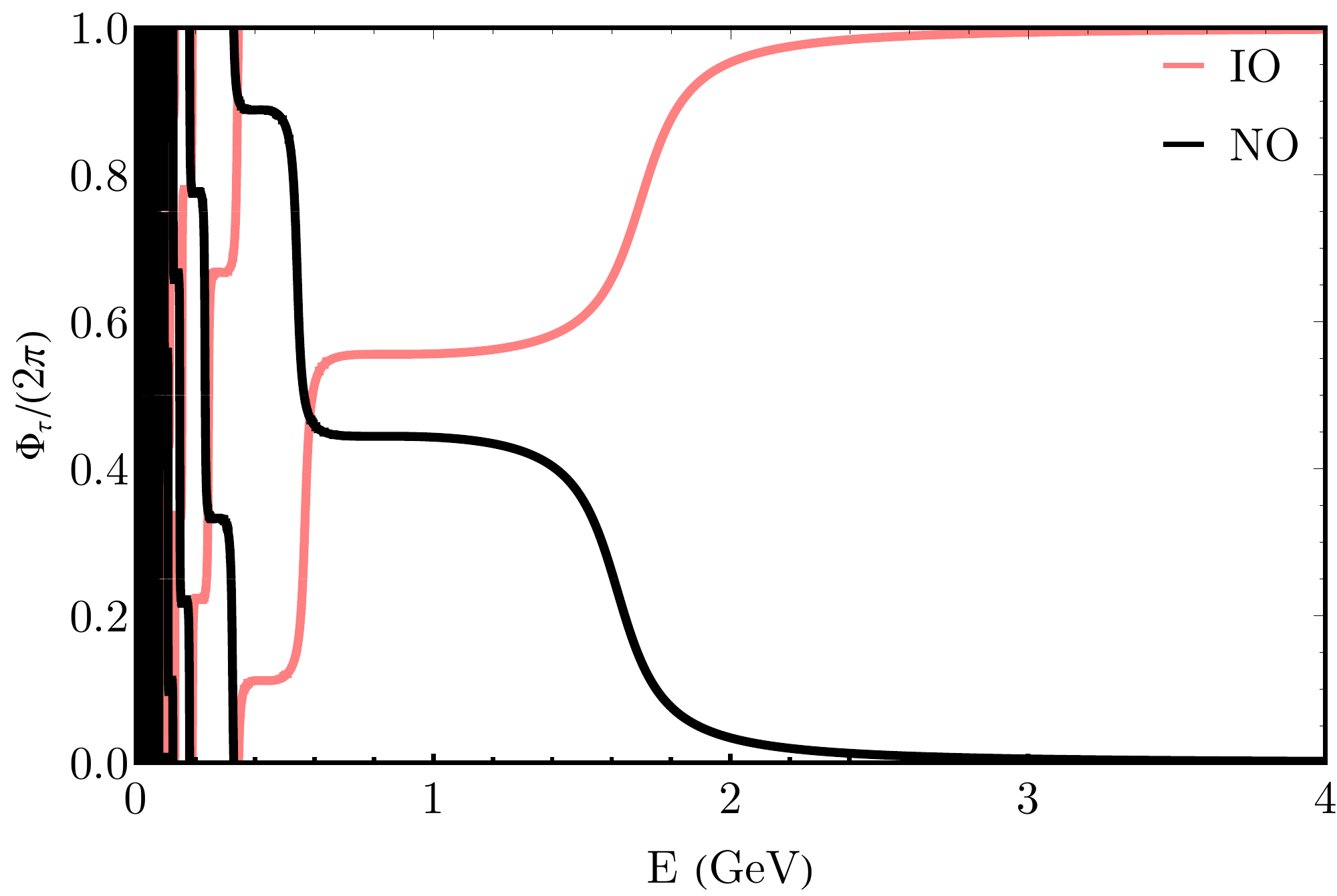}
	\caption{\label{fig:1}(colour online) Diagonal geometric phase $\Phi_{\alpha}$ as a function of $E$ for NO and IO.}
\end{figure}

For normal ordering (NO), i.e. when $\Delta M^2_{31}$ is positive, and for inverted ordering (IO), i.e. when $\Delta M^2_{31}$ is negative, $\Phi_{\alpha}$ approaches either 0 or $2\pi$ for large values of $E$ depending on $\alpha$. $\Phi_{\alpha}$ shows oscillatory behaviour for small $E$ due to the periodicity of the argument of a complex number. These results are shown in FIG. (\ref{fig:1}).

\begin{figure}[t]
\includegraphics[width=0.425\textwidth]{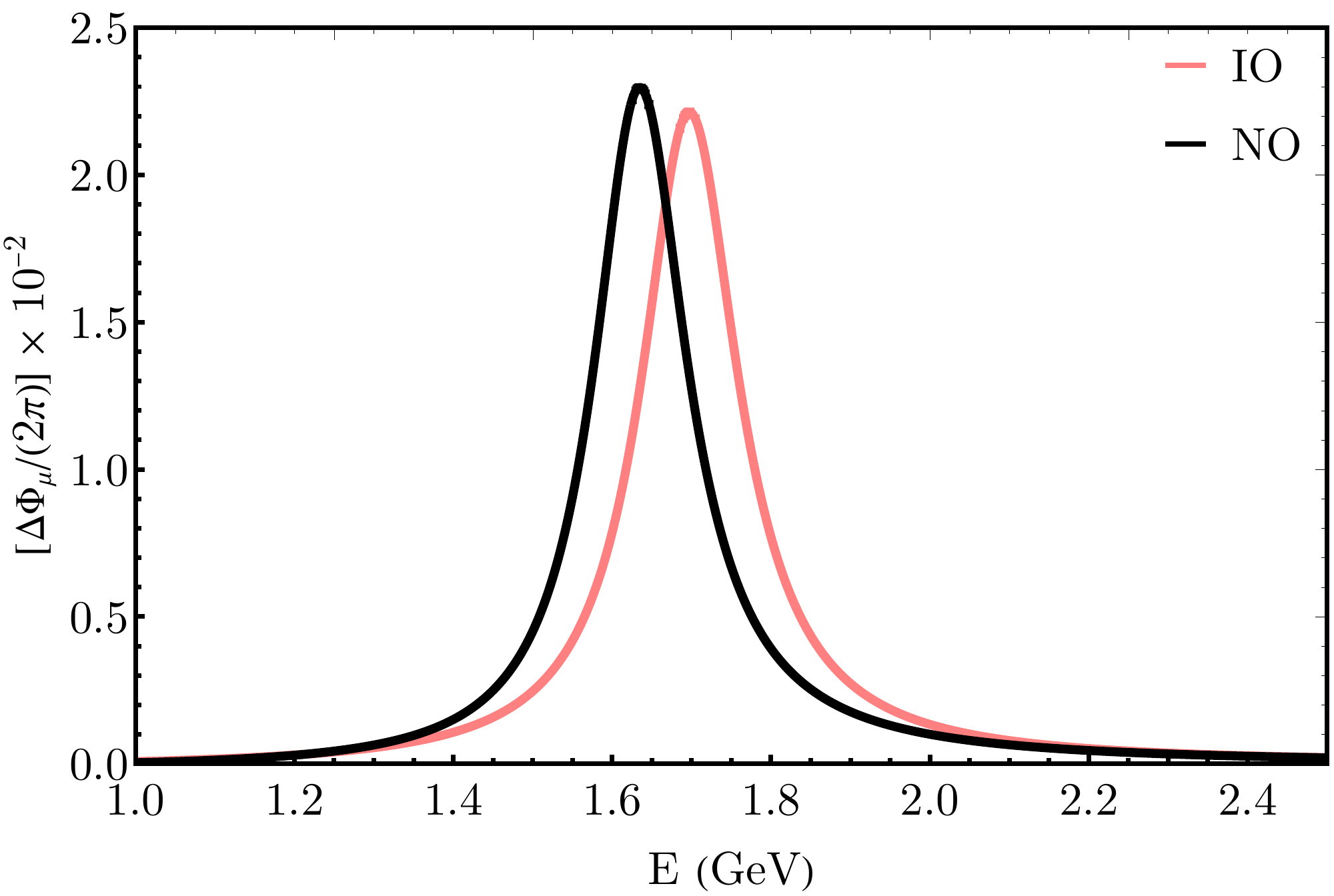}\\
\includegraphics[width=0.425\textwidth]{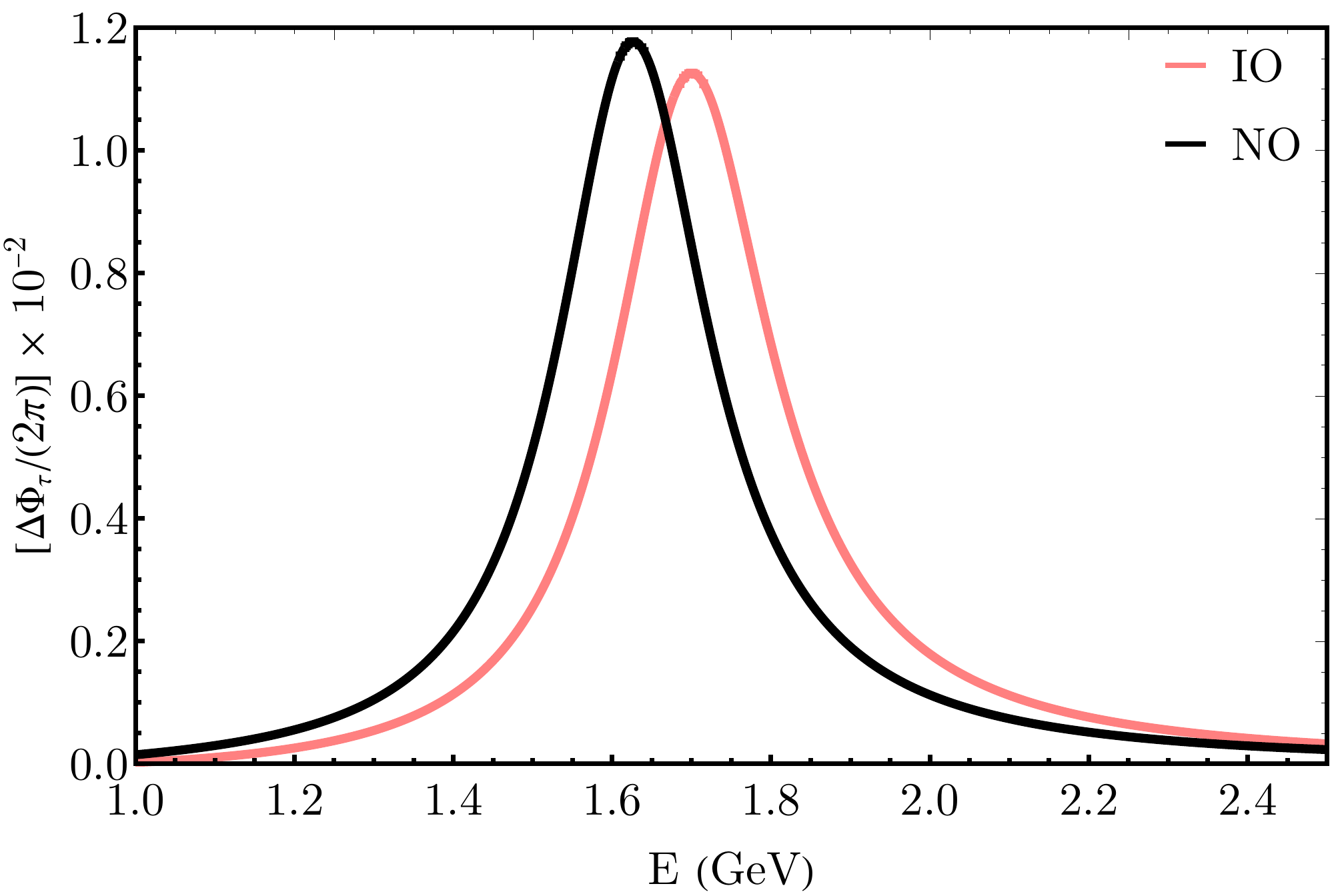}
	\caption{\label{fig:1a}(colour online) $\Delta\Phi_{\alpha}$ as function of $E$ for NO and IO, with $\delta=1.36\pi$ for $\delta\neq0$.}
\end{figure}

From the figures in FIG. (\ref{fig:1}), one might assume a symmetry between the curves for NO and IO about the value $\pi$. To show that they are not symmetric about $\pi$ and to highlight the sensitivity over the magnitude of $\delta$, we define an asymmetry factor of the form:
\begin{equation}
\Delta\Phi_{\alpha}=\Phi_{\alpha}(\delta\neq 0)-\Phi_{\alpha}(\delta=0).
\label{dgpaf}
\end{equation} 
The asymmetry factor in Eq. (\ref{dgpaf}) computes the difference between the diagonal geometric phases for different values of $\delta$ with a fixed mass ordering. The non-zero value of $\Delta\Phi_{\alpha}$ indicates the sensitivity of $\Phi_{\alpha}$ over $\delta$ and the shift between $\Delta\Phi_{\alpha}$ for different mass ordering indicates the asymmetry between $\Phi_{\alpha}$, which indicates the sensitivity over mass ordering. One can then interpret $\Delta\Phi_{\alpha}$ as a measure of sensitivity of the diagonal geometric phase over mass ordering and Dirac CP violating phase. We can see these features from FIG. (\ref{fig:1a}).

\subsection{Second order off-diagonal geometric phases in neutrinos}

Since we took the complete three flavour model, and there was no bi-maximal mixing (i.e. $\theta_{ij}\neq\pi/4$), we could not see any complete flavour transition. Hence, all $\Phi_{\alpha}$ were well defined in the energy range used to plot. However, that will not be the case if there was bi-maximal mixing, which we will address while discussing the MSW resonance effect in Sec. (\ref{Sec4D}). Also, we know that the diagonal geometric phases do not account for all the details of geometric phase. Therefore it is necessary to evaluate off-diagonal geometric phases.

In the three flavour neutrino model, there are three unique second order off-diagonal geometric phases namely, $\Phi_{\mu e}$, $\Phi_{\tau e}$ and $\Phi_{\tau\mu}$. For any distinct pair of flavours $\alpha$ and $\beta$, we can define
\begin{eqnarray}
\Phi_{\alpha\beta}=\arg\left[\sigma_{\alpha\beta}\sigma_{\beta\alpha}\right].
\end{eqnarray}
Then, using Eq. (\ref{sigmaab}) with $\beta$ as the initial flavour, we have,
\begin{align}
\sigma_{\alpha\beta}
=&\braket{\nu_{\alpha}|\nu_{\beta}(L)}\exp\left[-\int_{0}^{L}dL'\bra{\nu_{\beta}(L')}\partial_{L'}\ket{\nu_{\beta}(L')}\right].
\end{align}
Here the inner product becomes,
\begin{align}
\braket{\nu_{\alpha}|\nu_{\beta}(L)}= &U_{\alpha1} {U_{\beta1}^*} +   U_{\alpha2}{U_{\beta2}^*} \exp\left\{- \frac{i \Delta{M^2_{21}} L}{E}\right\}\nonumber\\&+U_{\alpha3}U_{\beta3}^*  \exp\left\{- \frac{i \Delta{M^2_{31}} L}{E}\right\}.
\end{align}
Then using Eq. (\ref{dynamicaloffset}) with $\beta$ as the flavour index, we get,
\begin{align}
\sigma_{\alpha\beta}= &\left(U_{\alpha1} {U_{\beta1}^*} +   U_{\alpha2}{U_{\beta2}^*} \exp\left\{- \frac{i \Delta{M^2_{21}} L}{E}\right\}\right.\nonumber\\&+\left.U_{\alpha3}U_{\beta3}^*  \exp\left\{- \frac{i \Delta{M^2_{31}} L}{E}\right\}\right)\nonumber \\&\times\exp\left[i\frac{L}{E} \left( |U_{\beta 2}|^2 \Delta{M^2_{21}} + |U_{\beta 3}|^2 \Delta{M^2_{31}} \right)\right].
\label{sigmaba}
\end{align}
Using the above general expression one can then compute $\sigma_{\beta\alpha}$ and the expression for $\Phi_{\alpha\beta}$ takes the form,
\begin{align}
\Phi_{\alpha\beta}=&\arg\left[\left(U_{\alpha1} {U_{\beta1}^*} +   U_{\alpha2}{U_{\beta2}^*} \exp\left\{- \frac{i \Delta{M^2_{21}} L}{E}\right\}\right.\right.\nonumber\\&+\left.U_{\alpha3}U_{\beta3}^*  \exp\left\{- \frac{i \Delta{M^2_{31}} L}{E}\right\}\right)\times\nonumber\\
&\left(U_{\beta1} {U_{\alpha1}^*} +   U_{\beta2}{U_{\alpha2}^*} \exp\left\{- \frac{i \Delta{M^2_{21}} L}{E}\right\}\right.\nonumber\\&+\left.U_{\beta3}U_{\alpha3}^*  \exp\left\{- \frac{i \Delta{M^2_{31}} L}{E}\right\}\right)\nonumber
\\&\times\exp\left\{i\frac{L}{E} \left[ \left(|U_{\beta 2}|^2+|U_{\alpha 2}|^2\right) \Delta{M^2_{21}}\right.\right. \nonumber\\&\left.+ \left(|U_{\beta 3}|^2+|U_{\alpha 3}|^2\right) \Delta{M^2_{31}} \right]\bigg\}\Bigg].
\label{phiab}
\end{align}
\begin{figure}[t]
	\includegraphics[width=0.425\textwidth]{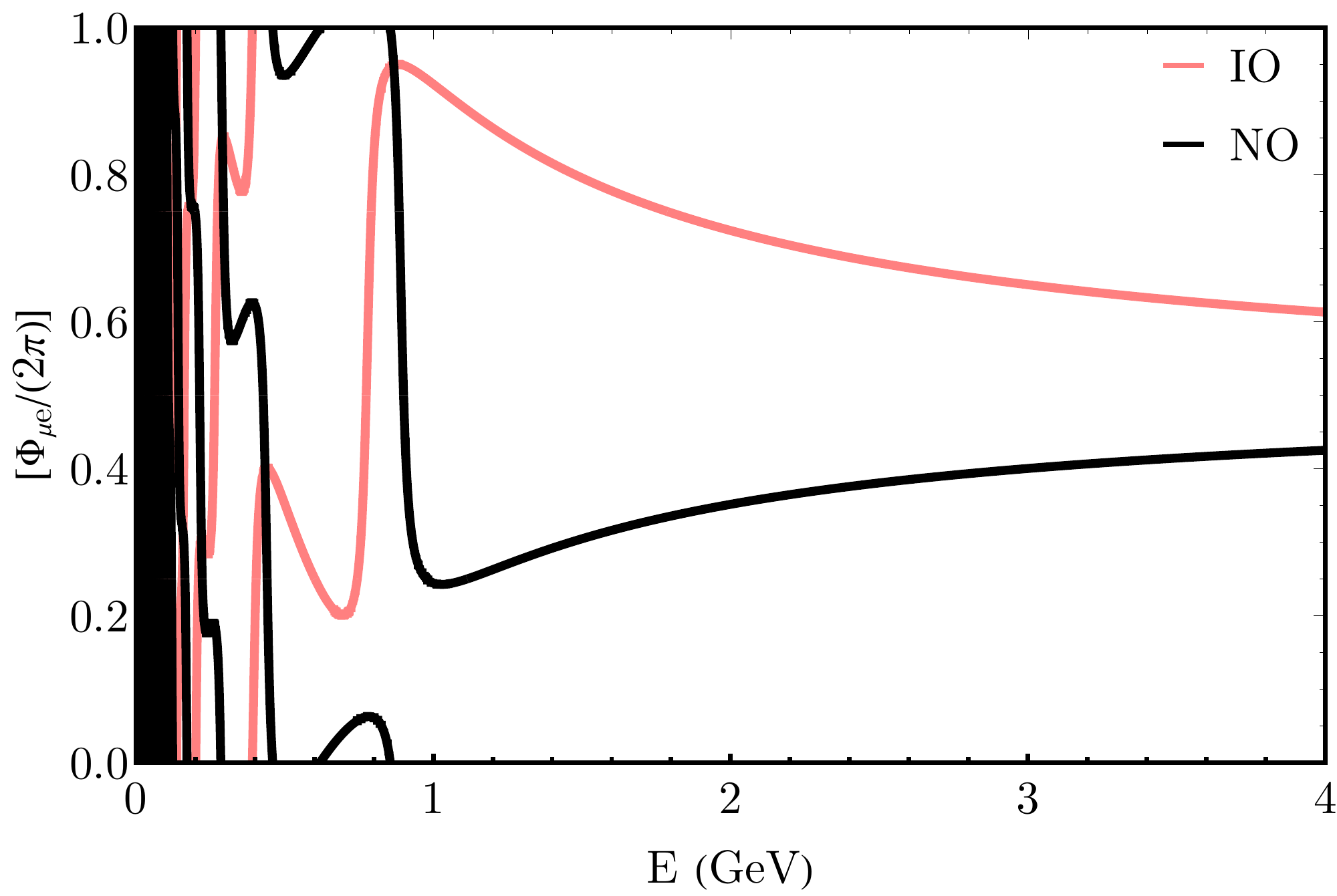}\\
	\includegraphics[width=0.425\textwidth]{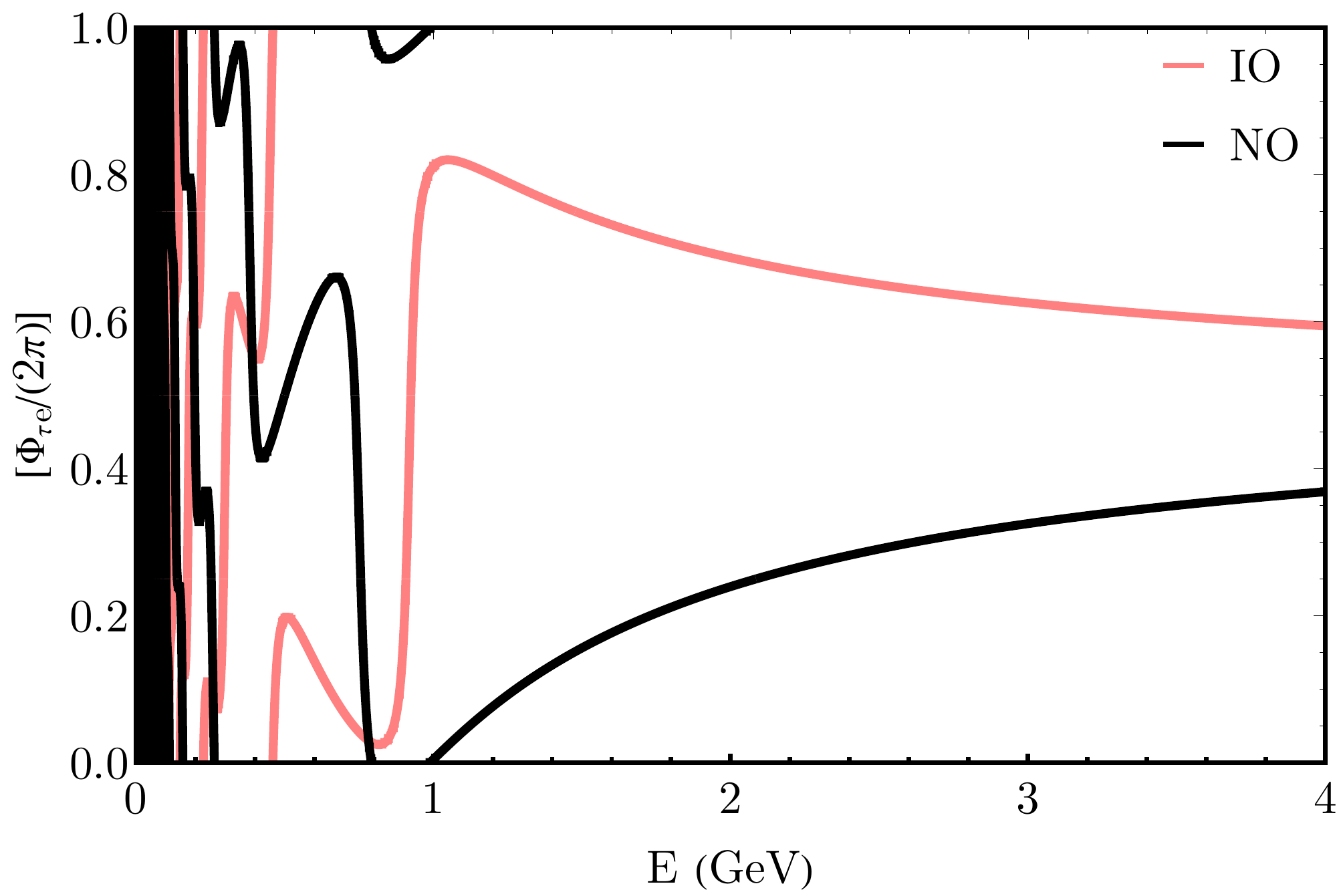}\\
	\includegraphics[width=0.425\textwidth]{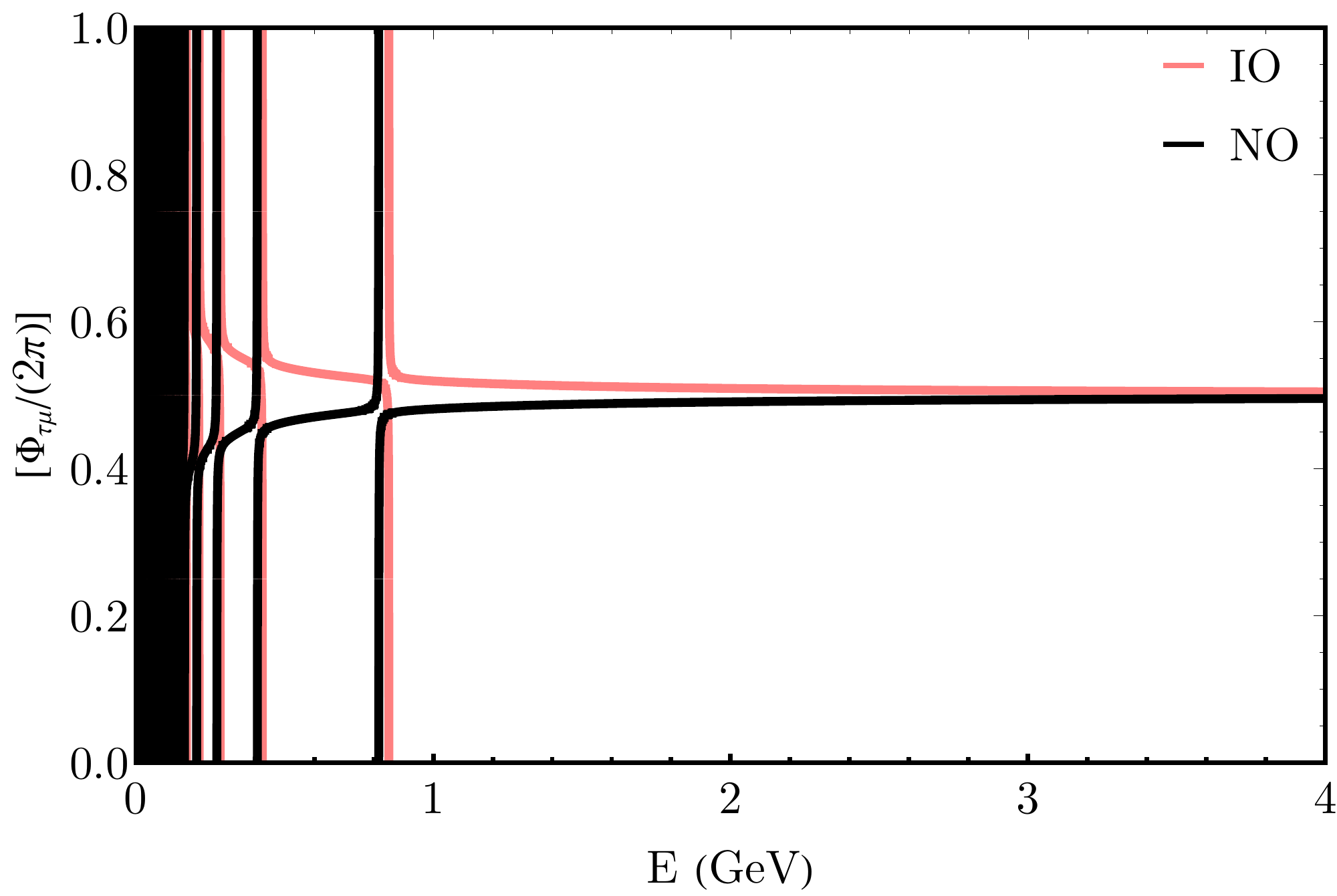}
	\caption{\label{fig:2}(colour online) Second order off-diagonal geometric phase $\Phi_{\alpha\beta}$ as function of $E$ for NO and IO.}
\end{figure}
One can now analyse the features of all three $\Phi_{\alpha\beta}$ by plotting them as a function of $E$ for a fixed $L$ as before. From FIG. (\ref{fig:2}) it is evident that all three $\Phi_{\alpha\beta}$ approaches $\pi$ for very large $E$ irrespective of the mass ordering and there is a huge asymmetry between NO and IO. Except for $\Phi_{\tau\mu}$, both $\Phi_{\mu e}$ and $\Phi_{\tau e}$ have higher sensitivity over mass ordering and $\delta$ in the energy range used. However, similar to diagonal geometric phases, the second order off-diagonal geometric phases are only sensitive to the magnitude of $\delta$. 

To show this let us take $\sigma_{\alpha\beta}
$ as a function of $\delta$ alone by fixing all other parameters. Clearly, from Eq. (\ref{phiab}), the terms inside the exponential function is not sensitive to the sign of $\delta$, as all $|U_{\alpha i}|^2$ either depend on $\cos\delta$ or are independent of $\delta$. If there is any dependency on the sign of $\delta$, then it must come from $\braket{\nu_{\alpha}|\nu_{\beta}(L)}\braket{\nu_{\beta}|\nu_{\alpha}(L)}$. In the expressions of all $U_{\alpha i}$ the dependency on $\delta$ comes in the form of $\exp(\pm i\delta)$. Then a little algebra will show that,
\begin{align}
\braket{\nu_{\alpha}|\nu_{\beta}(L)}=&g+he^{i\delta},\\
\braket{\nu_{\beta}|\nu_{\alpha}(L)}=&g+he^{-i\delta}.
\end{align}
Here, $g$ and $h$ are functions of all other mixing and oscillation parameters except $\delta$. Then,
\begin{align}
\braket{\nu_{\alpha}|\nu_{\beta}(L)}\braket{\nu_{\beta}|\nu_{\alpha}(L)}=&g^2+gh\left(e^{i\delta}+e^{-i\delta}\right)+h^2\nonumber\\
=&g^2+2gh\cos\delta+h^2.
\end{align}
Since $\cos\delta$ is an even function, the sensitivity on the sign of $\delta$ is lost for all $\Phi_{\alpha\beta}$. 
\begin{figure}[t]	\includegraphics[width=0.425\textwidth]{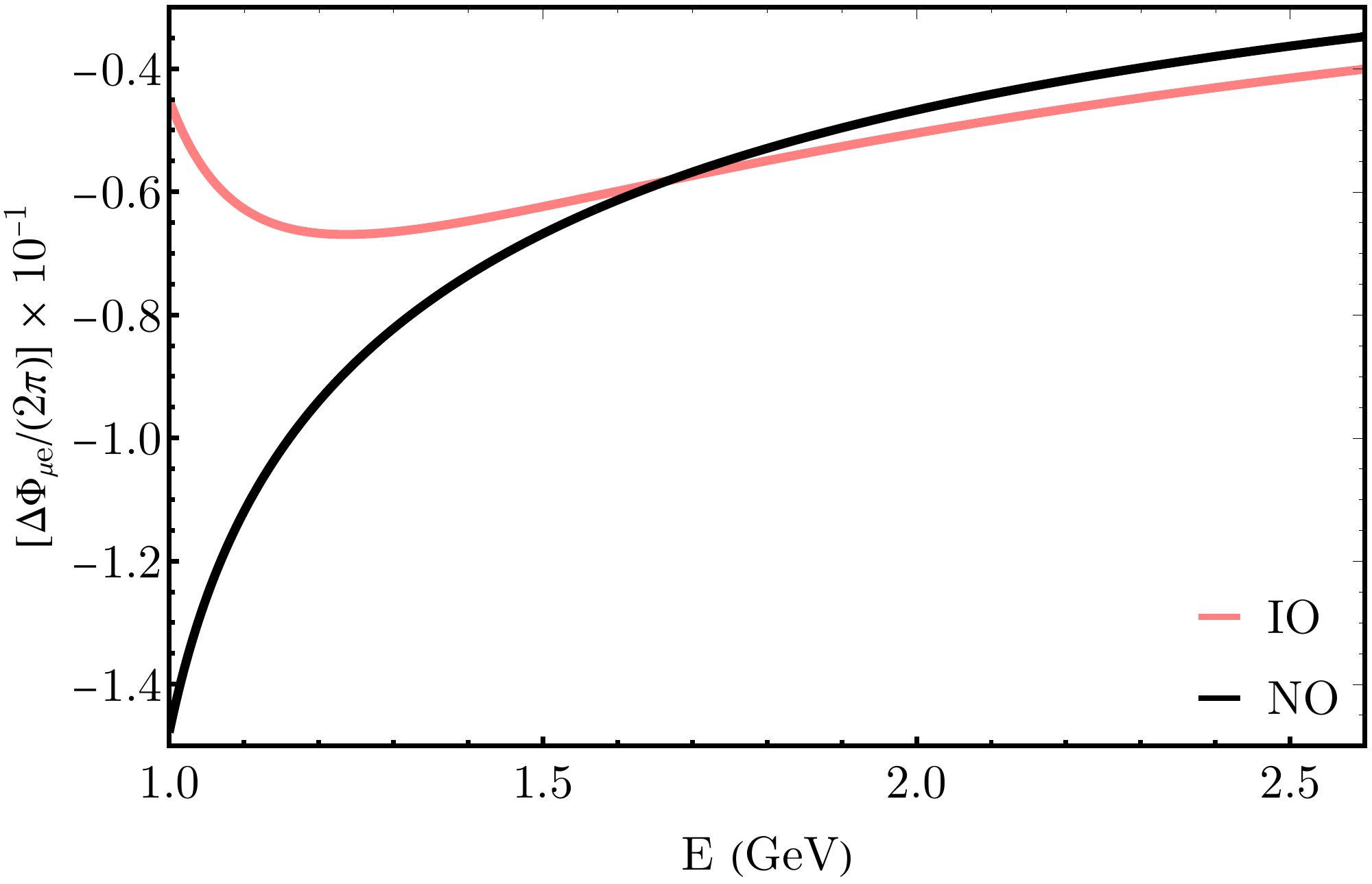}\\
	\includegraphics[width=0.425\textwidth]{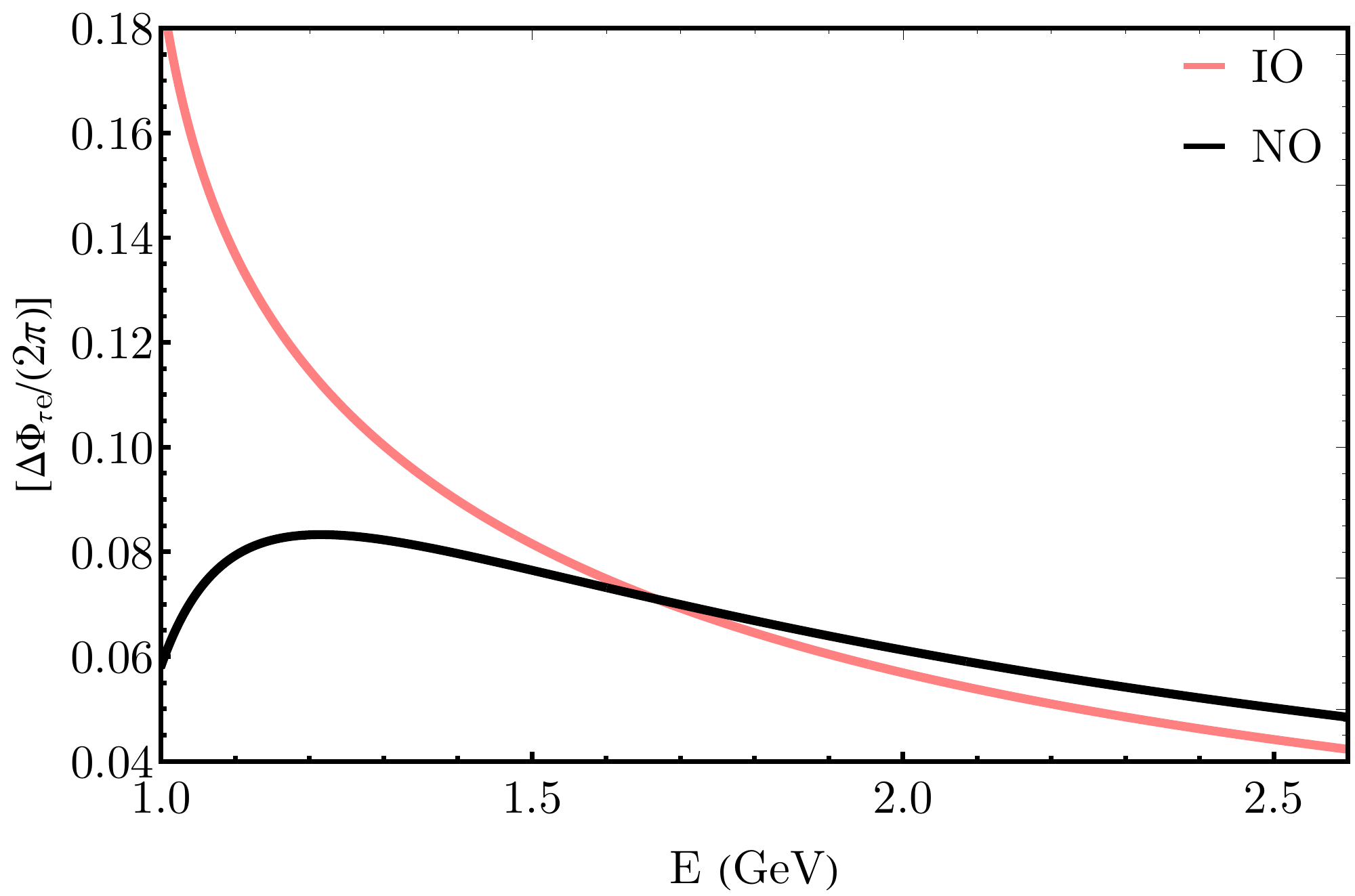}\\
	\includegraphics[width=0.425\textwidth]{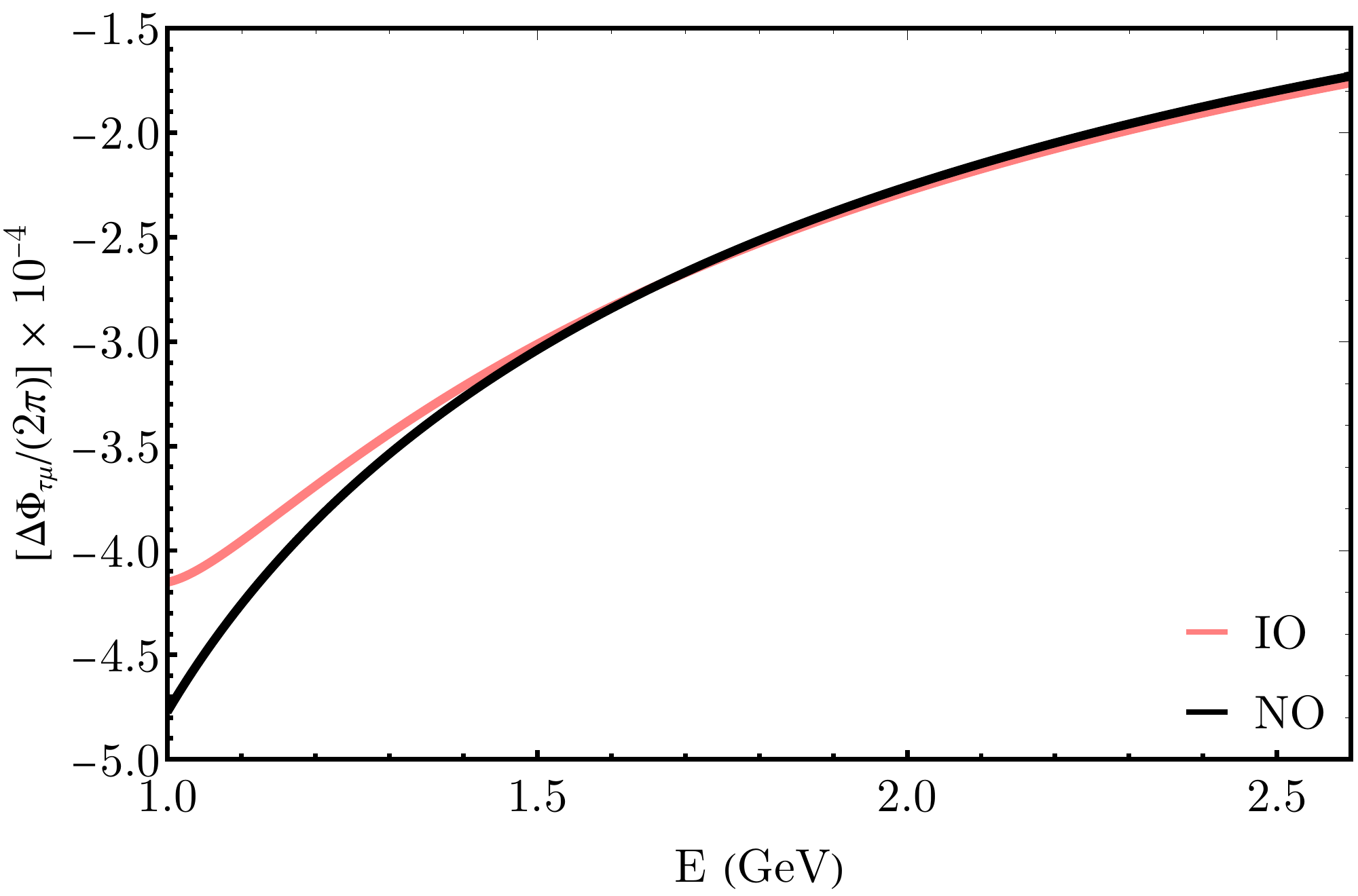}
	\caption{\label{fig:2a}(colour online) $\Delta\Phi_{\alpha\beta}$ as function of $E$ for NO and IO, with $\delta=1.36\pi$ for $\delta\neq0$.}
\end{figure}

Similar to the asymmetry factor for diagonal geometric phase, we define the asymmetry factor for second order off-diagonal geometric phase as, 
\begin{equation}
\Delta\Phi_{\alpha\beta}=\Phi_{\alpha\beta}(\delta\neq0)-\Phi_{\alpha\beta}(\delta=0)
\end{equation}

A non-zero value for $\Delta\Phi_{\alpha\beta}$ represents the sensitivity over the Dirac CP violating phase $\delta$ and the shift between $\Delta\Phi_{\alpha\beta}$, for each mass ordering, accounts the sensitivity over mass ordering. Plots in FIG. (\ref{fig:2}) and FIG.~(\ref{fig:2a}) illustrate the features of $\Phi_{\alpha\beta}$ and $\Delta\Phi_{\alpha\beta}$ respectively as a function of $E$ for each pairs of $\alpha$ and $\beta$. Clearly, second order off-diagonal geometric phases $\Phi_{\tau e}$ and $\Phi_{\mu e}$ are more sensitive to mass ordering and magnitude of delta $\delta$ than any diagonal geometric phases in the given energy range. While, $\Phi_{\tau\mu}$ shows the least sensitivity. Till now, all the gauge invariant geometric phases defined were sensitive to the mass ordering and the magnitude of $\delta$.

\subsection{Third order off-diagonal geometric phases in neutrinos}

The highest order gauge invariant geometric quantities that one can define for an $N$ level system are the $N^{\text{th}}$ order off-diagonal geometric phases. In our case we have a three level system and the third order off-diagonal geometric phases are,
\begin{align}
\Phi_{\alpha\beta\gamma}=\arg\left[\sigma_{\alpha\beta}\sigma_{\beta\gamma}\sigma_{\gamma\alpha}\right]\\
\Phi_{\beta\alpha\gamma}=\arg\left[\sigma_{\beta\alpha}\sigma_{\alpha\gamma}\sigma_{\gamma\beta}\right]
\end{align} 
The above equations are invariant under all cyclic permutations of flavour indices. Now, using the expression given by Eq. (\ref{sigmaba}), we get,
\begin{align}
\sigma_{\alpha\beta}\sigma_{\beta\gamma}\sigma_{\gamma\alpha}=&\braket{\nu_{\alpha}(0)|\nu_{\beta}(L)}\braket{\nu_{\beta}(0)|\nu_{\gamma}(L)}\braket{\nu_{\gamma}(0)|\nu_{\alpha}(L)}\nonumber\\&\times\exp\left\{iL\left[\bra{\nu_{\gamma}(0)}\hat{H}_F\ket{\nu_{\gamma}(0)}\right.\right.\nonumber\\&+\bra{\nu_{\beta}(0)}\hat{H}_F\ket{\nu_{\beta}(0)}\nonumber\\&\left.\left.+\bra{\nu_{\alpha}(0)}\hat{H}_F\ket{\nu_{\alpha}(0)}\right]\right\}
\end{align}
Due to the unitarity of the PMNS matrix, we have,
\begin{equation}
\sum_{\alpha}|U_{\alpha1}|^2=\sum_{\alpha}|U_{\alpha2}|^2=\sum_{\alpha}|U_{\alpha3}|^2=1.
\end{equation}
Then the exponential factor reduces to 
\begin{align}
&\exp\left\{iL\left[\bra{\nu_{\gamma}(0)}\hat{H}_F\ket{\nu_{\gamma}(0)}\right.\right.+\bra{\nu_{\beta}(0)}\hat{H}_F\ket{\nu_{\beta}(0)}+\nonumber\\&\left.\left.\bra{\nu_{\alpha}(0)}\hat{H}_F\ket{\nu_{\alpha}(0)}\right]\right\}=
\exp\left\{i\frac{L}{E} \left[ \Delta{M^2_{21}}+\Delta{M^2_{31}} \right]\right\}
\end{align}
Now, the exponential factor is devoid of all PMNS matrix mixing parameters and depends on the sign of $\Delta M^2_{31}$. Thus the sensitivity over $\delta$ must arise from the product $\braket{\nu_{\alpha}|\nu_{\beta}(L)}\braket{\nu_{\beta}|\nu_{\gamma}(L)}\braket{\nu_{\gamma}|\nu_{\alpha}(L)}$. 
For instance, let us take $\alpha=\mu$, $\beta=e$ and $\gamma=\tau$. Then, we can show that,
\begin{align}
\braket{\nu_{\mu}|\nu_{e}(L)}=\,&r_1+r_2e^{i\delta}\\
\braket{\nu_{e}|\nu_{\tau}(L)}=\,&r_3+r_4e^{-i\delta}\\
\braket{\nu_{\tau}|\nu_{\mu}(L)}=\,&r_5+r_6e^{i\delta}+r_7e^{-i\delta}
\end{align}
Where, $r_1,r_2,r_3,r_4,r_5,r_6$ and $r_7$ are functions of mixing and oscillation parameters excluding $\delta$. Then, 
\begin{align}
&\braket{\nu_{\mu}|\nu_{e}(L)}\braket{\nu_{e}|\nu_{\tau}(L)}\braket{\nu_{\tau}|\nu_{\mu}(L)}= r_1 r_3 r_5 + r_1 r_4 r_6 + \nonumber\\&~~~r_1 r_4 r_7\left(e^{ -2i \delta}\right) + r_2 r_3 r_6 \left(e^{2 i \delta}\right) + r_2 r_3 r_7 + r_2 r_4 r_5 +\nonumber\\&~~ \left(r_1 r_3 r_6 + r_2 r_3 r_5 + r_2 r_4 r_6\right) e^{ i \delta} + \nonumber\\&~~ \left(r_1 r_3 r_7 + r_1 r_4 r_5 + r_2 r_4 r_7\right)e^{- i \delta}
\end{align}
On simplifying the expression by substituting the exact formulae for $r_1,r_2,r_3,r_4,r_5,r_6$ and $r_7$ we get, $r_1r_4r_7-r_2r_3r_6=0$. This implies,
\begin{align}
r_1 r_4 r_7\left(e^{ -2i \delta}\right) + r_2 r_3 r_6 \left(e^{2 i \delta}\right)&=2r_1 r_4 r_7\cos(2\delta)
\end{align}
Which is only sensitive to the magnitude of $\delta$. On the other hand, we have,
\begin{align}
&\left(r_1 r_3 r_6 + r_2 r_3 r_5 + r_2 r_4 r_6\right) - \left(r_1 r_3 r_7 + r_1 r_4 r_5 + r_2 r_4 r_7\right)=\nonumber\\&\left(e^{\frac{i \Delta{M^2_{21}} L}{E}} - 1\right)e^{- \frac{2 i L \left(\Delta{M^2_{21}} + \Delta{M^2_{31}}\right)}{E}}\times\nonumber\\& \left[e^{\frac{i \Delta{M^2_{21}} L}{E}} + e^{\frac{2 i \Delta{M^2_{31}} L}{E}} - e^{\frac{i \Delta{M^2_{31}} L}{E}} - e^{\frac{i L \left(\Delta{M^2_{21}} + \Delta{M^2_{31}}\right)}{E}}\right]\times\nonumber\\& \left(\sin{\theta_{12}} \sin{\theta_{13}} \sin{\theta_{23}} \cos{\theta_{12}} \cos^{2}{\theta_{13}} \cos{\theta_{23}}\right).
\end{align}
This implies, the terms $\left(r_1 r_3 r_6 + r_2 r_3 r_5 + r_2 r_4 r_6\right) e^{ i \delta}$ and $ \left(r_1 r_3 r_7 + r_1 r_4 r_5 + r_2 r_4 r_7\right)e^{- i \delta}$ will contribute elements with $\cos\delta$ and $\sin\delta$ making it sensitive to both magnitude and sign of $\delta$. Thus the product $\braket{\nu_{\alpha}|\nu_{\beta}(L)}\braket{\nu_{\beta}|\nu_{\gamma}(L)}\braket{\nu_{\gamma}|\nu_{\alpha}(L)}$ is sensitive to the magnitude and sign of $\delta$. 

Hence, $\sigma_{\alpha\beta}\sigma_{\beta\gamma}\sigma_{\gamma\alpha}$ is sensitive to both mass ordering and the Dirac CP violating phase. Now, the third order off-diagonal geometric phase is
\begin{align}
\Phi_{\alpha\beta\gamma}=&\arg\left[\left(U_{\alpha1} {U_{\beta1}^*} +   U_{\alpha2}{U_{\beta2}^*} \exp\left\{- \frac{i \Delta{M^2_{21}} L}{E}\right\}\right.\right.\nonumber\\&+\left.U_{\alpha3}U_{\beta3}^*  \exp\left\{- \frac{i \Delta{M^2_{31}} L}{E}\right\}\right)\times\nonumber\\
&\left(U_{\beta1} {U_{\gamma1}^*} +   U_{\beta2}{U_{\gamma2}^*} \exp\left\{- \frac{i \Delta{M^2_{21}} L}{E}\right\}\right.\nonumber\\&+\left.U_{\beta3}U_{\gamma3}^*  \exp\left\{- \frac{i \Delta{M^2_{31}} L}{E}\right\}\right)\times\nonumber\\
&\left(U_{\gamma1} {U_{\alpha1}^*} +   U_{\gamma2}{U_{\alpha2}^*} \exp\left\{- \frac{i \Delta{M^2_{21}} L}{E}\right\}\right.\nonumber\\&+\left.U_{\gamma3}U_{\alpha3}^*  \exp\left\{- \frac{i \Delta{M^2_{31}} L}{E}\right\}\right)\nonumber
\\
&\times\left.\exp\left\{i\frac{L}{E} \left[ \Delta{M^2_{21}}+\Delta{M^2_{31}} \right]\right\}\right].
\label{phiabg}
\end{align}
Hence, $\Phi_{\alpha\beta\gamma}$ is a gauge invariant quantity sensitive to both mass ordering and the Dirac CP violating phase.

Furthermore, the non-cyclic permutation of flavour indices create an additional off-diagonal geometric phase $\Phi_{\beta\alpha\gamma}$. Interestingly one can easily identify that,
\begin{equation}
\Phi_{\alpha\beta\gamma}(\delta)=\Phi_{\beta\alpha\gamma}(-\delta)
\label{toodpsymm}
\end{equation}
provided all other parameters are kept same. Thus, when $\delta=0$, we have $\Phi_{\alpha\beta\gamma}=\Phi_{\beta\alpha\gamma}$. Hence, non-cyclic permutation of indices gives different results only when there is CP violation or time reversal symmetry violation.  

\begin{figure}[t]
\includegraphics[width=0.425\textwidth]{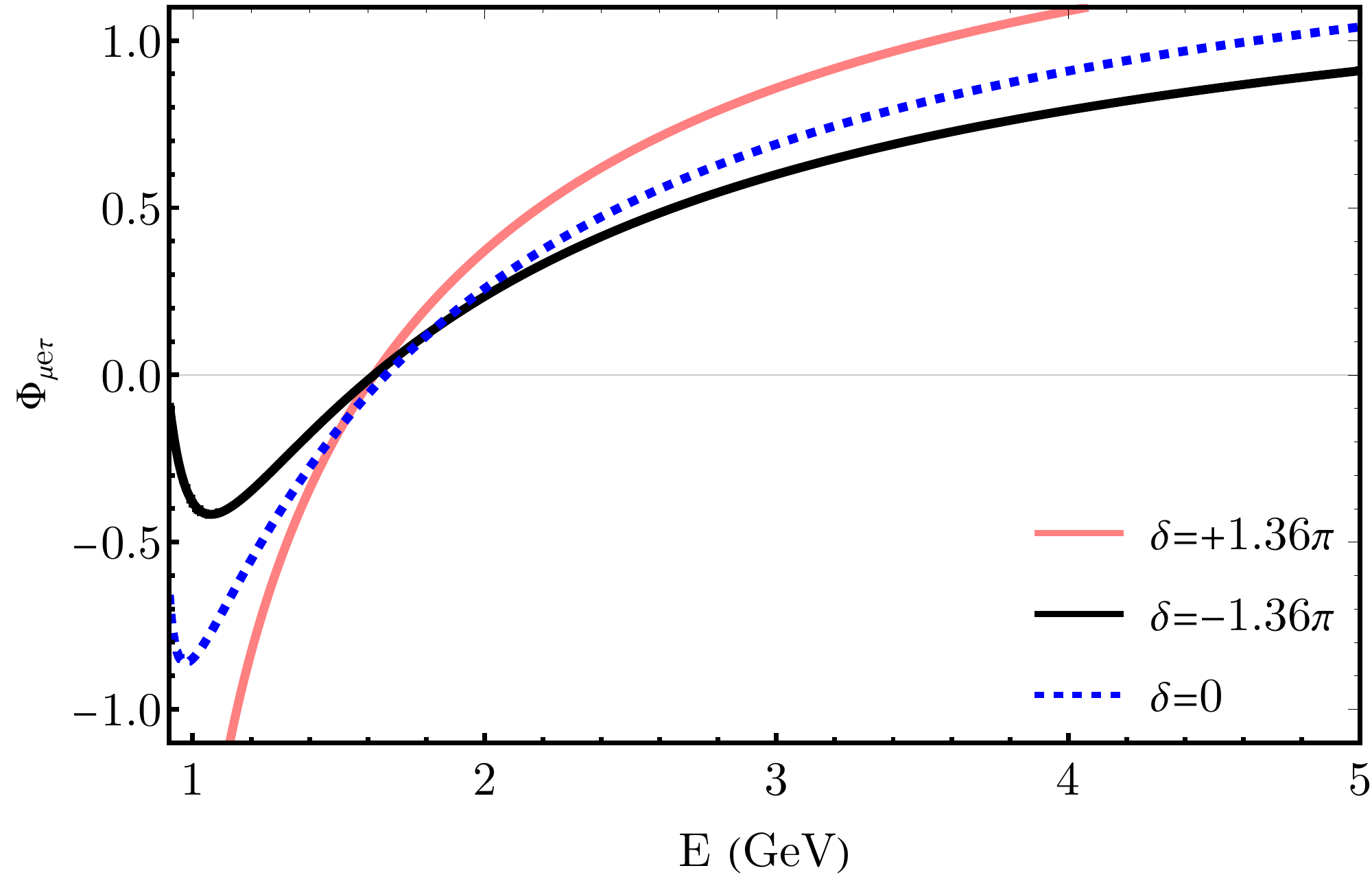}\\
\includegraphics[width=0.425\textwidth]{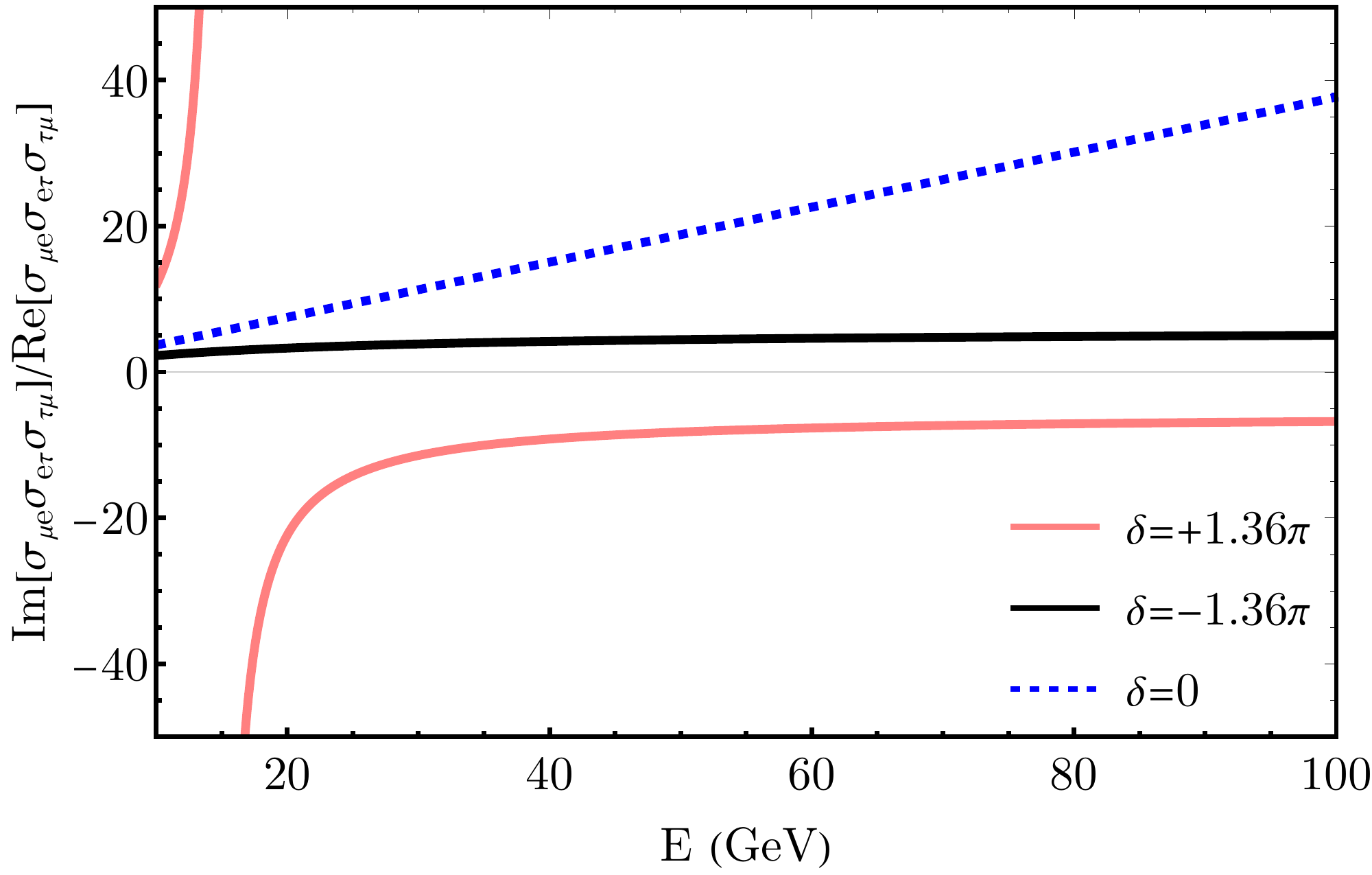}
	\caption{\label{fig:5}(colour online) $\Phi_{\mu e\tau}$ and the ratio between the imaginary and real part of $\sigma_{\mu e}\sigma_{e\tau}\sigma_{\tau\mu}$ as a function of $E$ for different values of $\delta$ with NO. Here the phases are defined between $-\pi$ and $\pi$.}
\end{figure}

For the clarity of illustration, here we define the geometric phase between $-\pi$ and $\pi$. For large $E$ with NO (IO), $\Phi_{\mu e \tau}$ converges to $\pi/2$ ($-\pi/2$  ) when $\delta=0$. For non-zero $\delta$, it converges to a value above or below $\pi/2$ depending on the sign of $\delta$. This implies, for $\delta=0$, the ratio between $\Im(\sigma_{\alpha\beta}\sigma_{\beta\gamma}\sigma_{\gamma\alpha})$ and $\Re(\sigma_{\alpha\beta}\sigma_{\beta\gamma}\sigma_{\gamma\alpha})$ keeps on increasing. While for $\delta\neq0$, it converges to a value above or below zero. For NO, if $\delta=\text{+ve}$,  $\Phi_{\mu e\tau}>\pi/2$ for very large $E$, thus the ratio diverges at a specific point. In our numerical study, this point is between $10$ and $20$ GeV. For IO with a positive $\delta$ we get $\Phi_{\mu e\tau}<-\pi/2$. The situation is reversed for  $\Phi_{\tau e\mu}$ based on Eq. (\ref{toodpsymm}). These properties are evident from FIG. (\ref{fig:5}).

Hence, the third order off-diagonal geometric phase is the only measurable geometric phase which is sensitive to the mass ordering and the Dirac CP violating phase. While all other lower order geometric phases are only sensitive to the mass ordering and the magnitude of $\delta$. In summary all diagonal and off-diagonal geometric phase gives the complete account of the geometric phases in neutrino mixing. A direct measurement of the same can address the existing open problems in a novel way.

\subsection{\label{Sec4D}Two flavour model and matter potential}

So far, we considered the case when $V_{CC}=0$. To include the effects of matter potential, it is convenient to study geometric phases using the two flavour neutrino model. One can approximate the three flavour model to a two flavour model depending on the parameters $E$ and $L$. In doing so, we lose the sensitivity over mass ordering and CP violation. However, the approximation is helpful to include the effects of matter potential and thus recover the sensitivity over mass ordering. 

In the case of two flavour oscillations, we can have two situations where $L/E$ is either small or large compared to $1/\Delta m^2$. Under these situation we look for either $\ket{\nu_{\mu}}\rightarrow\ket{\nu_{\tau}}$ or $\ket{\nu_e}\rightarrow\ket{\nu_{\mu/\tau}}$ oscillations respectively. Irrespective of the flavours, one can assume these two specific flavour combinations as $\alpha$ and $\beta$ with mixing angle $\theta$ and mass squared difference $\Delta m^2$. Now, the PMNS matrix for two flavour model become,
\begin{align}
U&={\begin{pmatrix}U_{\alpha1}&U_{\alpha2}\\
	U_{\beta 1}&U_{\beta 2}
	\end{pmatrix}}={\begin{pmatrix}\cos\theta&\sin\theta\\
	-\sin\theta&\cos\theta
	\end{pmatrix}}
\end{align}
Following the same steps as in the case of three flavour model, the geometric phases factors are
\begin{align}
\sigma_{\alpha\alpha}=\braket{\nu_{\alpha}(0)|\nu_{\alpha}(L)}\exp\left[iL\bra{\nu_{\alpha}(0)}\hat{H}_F\ket{\nu_{\alpha}(0)}\right],\\
\sigma_{\alpha\beta}=\braket{\nu_{\alpha}(0)|\nu_{\beta}(L)}\exp\left[iL\bra{\nu_{\beta}(0)}\hat{H}_F\ket{\nu_{\beta}(0)}\right].
\end{align}
Then the two independent diagonal geometric phases are
	\begin{align}
	\Phi_{\alpha}=\arg&\left\{\left[\cos^{2}{\theta } + \exp\left\{- \frac{i \Delta{M^2} L}{E}\right\} \sin^{2}{\theta}\right]\right.\nonumber\\&\times\left.\exp\left[\frac{i \Delta{M^2} L \sin^{2}{\theta}}{E}\right]\right\}\\
	\Phi_{\beta}= \arg&\left\{\left[\sin^{2}{\theta} + \exp\left\{- \frac{i \Delta{M^2} L}{E}\right\} \cos^{2}{\theta}\right]\right.\nonumber\\&\times\left.\exp\left[\frac{i \Delta{M^2} L \cos^{2}{\theta}}{E}\right]\right\}
	\end{align}
	And the off-diagonal geometric phase is
	\begin{align}
	\Phi_{\alpha\beta}= \arg&\left\{ \left(1 - \exp\left\{\frac{i \Delta{M^2} L}{E}\right\}\right)^{2} \sin^2\theta\cos^2\theta \right.\nonumber\\&\times\left. \exp\left\{- \frac{ i \Delta{M^2} L}{E}\right\}\right\}
	\label{off2F}
	\end{align}
	
 For two flavour model, one can redefine $\theta$ and $\Delta m^2$ to $\theta_M$ and $\Delta m^2_M$ as given by Eq. (\ref{MSWangle}) and Eq. (\ref{MSWmass}) respectively, such that the equations derived later retains the same form. The factor $1-(2EV_{CC}/\Delta m^2\cos(2\theta))$ takes different values for positive and negative values of $\Delta m^2$ and becomes sensitive to mass ordering. Additionally,  $V_{CC}$ is positive for neutrinos and negative for anti-neutrinos. For $2EV_{CC}=\Delta m^2\cos(2\theta)$, known as Mikheyev-Smirnov-Wolfenstein (MSW) resonance, the mixing angle becomes $\pi/4$. The region with
\begin{equation}
V_{CC}=\frac{\Delta m^2\cos(2\theta)}{2E}
\label{MSWresonance}
\end{equation}
is what we call as the MSW resonance region. Thus for high energy neutrino, a low electron density region can show MSW effect and vice versa. In two flavour approximation, although we could study $\ket{\nu_{\mu}}\rightarrow\ket{\nu_{\tau}}$ oscillation, we need not consider a matter potential as we do not expect a muon or tau rich medium, except may be in cosmological epochs. Thus to investigate the current mass ordering problem associated with $\Delta m^2_{31}$, one must resort to the complete three flavour model. 

Here, we explore the situation with solar neutrino parameters by setting $\theta=\theta_{12}=33.64^{\circ}$ and $\Delta m^2=\Delta m^2_{21}=7.53\times10^{-5}$ eV$^2$. In our expressions, we have $\Delta M^2:=2.54\Delta m^2$ to avoid unnecessary numerical factors. For solar neutrinos oscillation, which is mostly, $\ket{\nu_{e}}\rightarrow\ket{\nu_{\mu/\tau}}$ the first oscillation maximum is approximately around the ratio $L/E\approx15,000$ km/GeV. For illustration purpose, we assume a distance of $L=15$~km with energy $E$ under $2$ MeV.

Here, the diagonal phase takes values either 0 or $\pi$ depending on the mass ordering in the MSW resonance region given by Eq. (\ref{MSWresonance}). Outside the resonance region, the diagonal geometric phases takes non-trivial values. In the MSW resonance region, the value of $\theta_M=\pi/4$ and $\Delta M^2_M=2.54\times\Delta m^2_M$. Then, 
\begin{align}
\Phi_{\alpha}=&\arg\left\{\left[\frac{1}{2}+\frac{1}{2}\exp\left\{-\frac{i\Delta M^2_ML}{E}\right\}\right]\exp\left[\frac{i\Delta M^2_ML}{2E}\right]\right\}\nonumber\\
=&\arg\left[\cos\left(\frac{\Delta M^2_ML}{2E}\right)\right].
\end{align}
Then,
\begin{align}
\Phi_{\alpha}=\left\{\begin{matrix}
\pi &\text{ for } \Delta M^2_ML/(2E)>\pi/2\\\\
0 &\text{ for } \Delta M^2_ML/(2E)<\pi/2\\\\
\text{undefined}&\text{ for }\Delta M^2_ML/(2E)=\pi/2
\end{matrix} \right..
\end{align}
Here, $\Delta M^2_ML/(2E)=\pi/2$ corresponds to the complete oscillation inversion at,
\begin{equation}
E=\frac{2.54\times\Delta m^2_M L}{\pi}.
\end{equation}
The factor $2.54$ comes due to the convention we follow (See Sec. (\ref{UC})).

\begin{figure}[h!]
	\subfloat{\includegraphics[width=1\columnwidth]{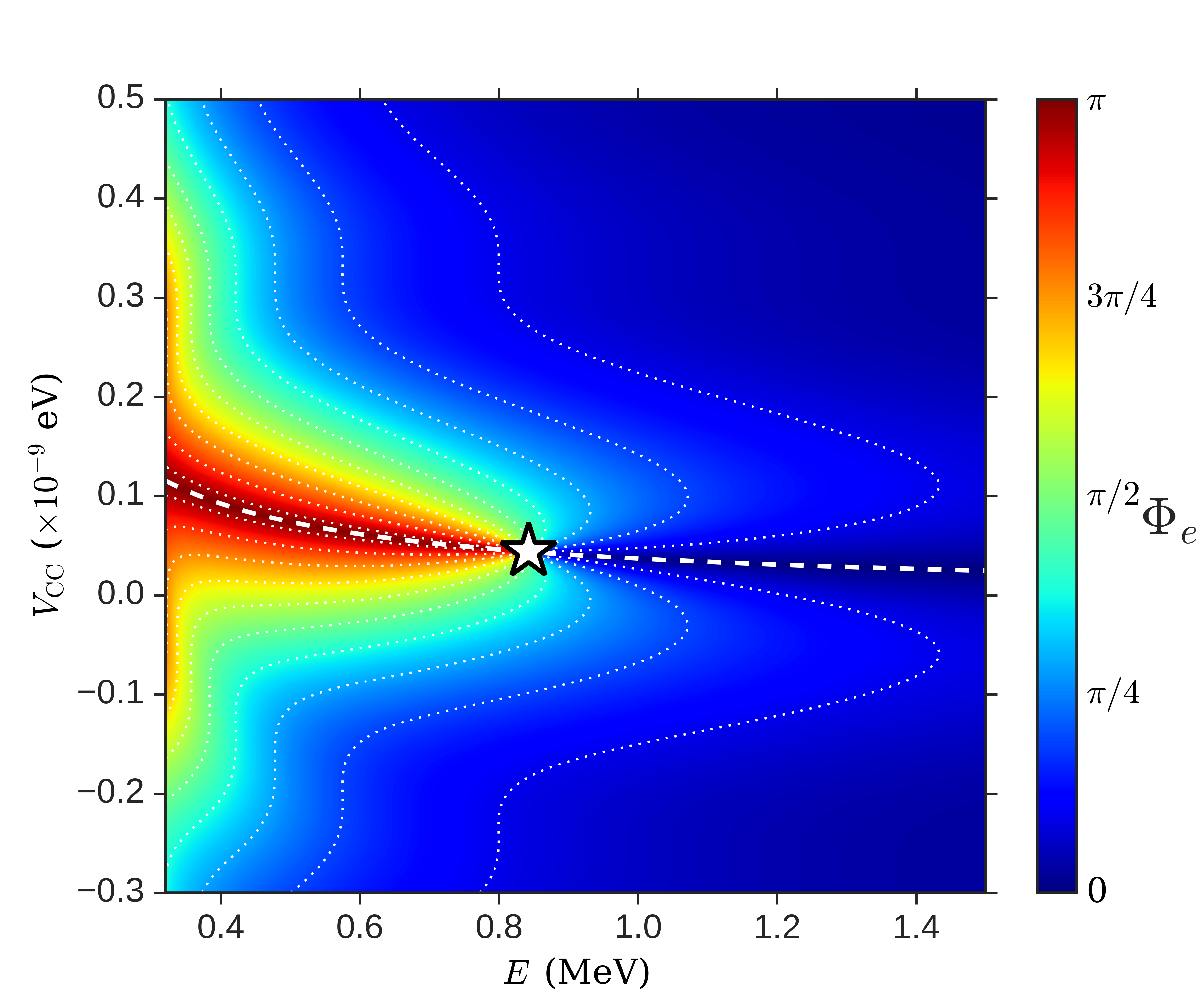}}
	\caption{(colour online) $\Phi_e$ as the function of $E$ and matter potential $V_{CC}$ for NO. The dashed white curve represents the MSW resonance region by Eq.~(\ref{MSWresonance}). The star denotes the nodal point.\label{MSWplot}}
\end{figure}

Considering the MSW resonance region, we have a sudden shift of $\pi$ at complete oscillation inversion. This jump is because, the final state is orthonormal to the initial state and the inner product between them is zero. 
The point is interesting as this corresponds to a nodal point, and here diagonal geometric phase becomes undefined. FIG. (\ref{MSWplot}) presents these features for normal ordering. 

An interesting thing to note is the correspondence between $\Phi_{\alpha}$ and $\Phi_{\beta}$, where one can replace all $\sin\theta$ with $\cos\theta$ and vice versa to get each other. Thus the transformation of the form $\theta\rightarrow\frac{\pi}{2}-\theta$ implies $\Phi_{\alpha}\rightarrow\Phi_{\beta}$. We can then say that one diagonal geometric phase is equal to the other evaluated at the complementary angle. Hence, $\Phi_{\alpha}$ is the co-function of $\Phi_{\beta}$ with respect to the mixing angle for a two flavour model. 

Now, considering the only off-diagonal geometric phase in the two flavour model given by Eq. (\ref{off2F}) we get,
\begin{align}
\Re\left[\sigma_{\alpha\beta}\right]&= -4\sin^2\left(\frac{\Delta M^2L}{2E}\right)\sin^2\theta\cos^2\theta\\
\Im\left[\sigma_{\alpha\beta}\right]&=0.
\end{align}
Then the real part is always negative, except for $E\rightarrow\infty$ and $L=0$, and the imaginary part of $\sigma_{\alpha\beta}$ is always zero. Thus $\Phi_{\alpha\beta}$ is always $\pi$ irrespective of any mixing parameters, and undefined for $E\rightarrow\infty$ and $L=0$. Hence, the only second order off-diagonal geometric phase in two flavour approximation is a topological invariant quantity. This calculation stands as an alternative proof for the results by P.~Mehta~\cite{PhysRevD.79.096013}. The geometric phases appearing in the neutrino mixing has a non trivial topological origin from the structure of the PMNS mixing formalism. Additionally, our solutions reduce to previous results \cite{BLASONE1999262,PhysRevD.63.053003} under appropriate conditions.
 
\section{\label{Sec.V}Conclusion}
To summarise, we studied the geometric phases arising in neutrino mixing using the kinematic approach. We derived the general formulae for all the plausible gauge invariant geometric phases in three and two flavour neutrino mixing models. Our results are, in principle, generalizable, as we did not restrict to any particular parametrization of the PMNS mixing matrix. The general expressions are suitable for the implementation of numerical analysis and thus extendable to a three flavour model including matter effects and other non-standard interactions.

We obtained the general expressions for all three diagonal geometric phases and prove that they are sensitive to the mass ordering and to the magnitude of the Dirac CP violating phase. Further, we derived the general formulae to compute all second order off-diagonal geometric phases. Again, these phases are sensitive to the mass ordering and the magnitude of the Dirac CP violating phase. Even though second order off-diagonal geometric phases cannot distinguish between the sign of $\delta$, they are more sensitive than the diagonal geometric phases. The asymmetry factor we defined accounts for the sensitivity over mass ordering and $\delta$ of different geometric phases. Finally, in the three flavour model, we derived the expressions for the third order off-diagonal geometric phase. We showed that it is sensitive to the mass ordering and the sign of $\delta$.
We found that the non-cyclic permutations of flavour indices in the third order off-diagonal geometric phase is immaterial only when $\delta=0$. For non-zero $\delta$, $\Phi_{\alpha\beta\gamma}(\delta)=\Phi_{\beta\alpha\gamma}(-\delta)$. 

To investigate the effects of matter potential, we considered the two flavour model and derived the expressions for diagonal and off-diagonal geometric phases. We show that the two possible diagonal geometric phases are co-functions with respect to the mixing angle and the only off-diagonal geometric is a topological invariant quantity which always assumes value $\pi$. The topological origin is rooted in the structure of PMNS mixing formalism. In the MSW resonance region, we illustrate that the diagonal phase is either 0 or $\pi$, and at complete oscillation inversion, it is undefined. The transition between zero and $\pi$ occurs at complete oscillation inversion, which is called the nodal point.

In short geometric phases opens an alternative way to look at neutrino physics and shows potential to resolve ongoing open problems such as mass ordering and estimation of the Dirac CP violating phase. Finally, one can make use of our results in any field with structures similar to oscillation physics.

\begin{acknowledgments}
MTM  sincerely  thank  C.  P.  Jisha,  Friedrich  Schiller Universit\"{a}t  Jena, for  the  fruitful  discussions  on  geometric  phases and acknowledges CSIR JRF, Government  of  India,  Grant No: 09 / 239(0558) / 2019-EMR-I for  the  financial  support.  Authors gratefully acknowledge the financial support by the Department of Science and Technology, India. All authors are thankful to the referees for their valuable comments.
\end{acknowledgments}


\begin{thebibliography}{58}%
	\makeatletter
	\providecommand \@ifxundefined [1]{%
		\@ifx{#1\undefined}
	}%
	\providecommand \@ifnum [1]{%
		\ifnum #1\expandafter \@firstoftwo
		\else \expandafter \@secondoftwo
		\fi
	}%
	\providecommand \@ifx [1]{%
		\ifx #1\expandafter \@firstoftwo
		\else \expandafter \@secondoftwo
		\fi
	}%
	\providecommand \natexlab [1]{#1}%
	\providecommand \enquote  [1]{``#1''}%
	\providecommand \bibnamefont  [1]{#1}%
	\providecommand \bibfnamefont [1]{#1}%
	\providecommand \citenamefont [1]{#1}%
	\providecommand \href@noop [0]{\@secondoftwo}%
	\providecommand \href [0]{\begingroup \@sanitize@url \@href}%
	\providecommand \@href[1]{\@@startlink{#1}\@@href}%
	\providecommand \@@href[1]{\endgroup#1\@@endlink}%
	\providecommand \@sanitize@url [0]{\catcode `\\12\catcode `\$12\catcode
		`\&12\catcode `\#12\catcode `\^12\catcode `\_12\catcode `\%12\relax}%
	\providecommand \@@startlink[1]{}%
	\providecommand \@@endlink[0]{}%
	\providecommand \url  [0]{\begingroup\@sanitize@url \@url }%
	\providecommand \@url [1]{\endgroup\@href {#1}{\urlprefix }}%
	\providecommand \urlprefix  [0]{URL }%
	\providecommand \Eprint [0]{\href }%
	\providecommand \doibase [0]{http://dx.doi.org/}%
	\providecommand \selectlanguage [0]{\@gobble}%
	\providecommand \bibinfo  [0]{\@secondoftwo}%
	\providecommand \bibfield  [0]{\@secondoftwo}%
	\providecommand \translation [1]{[#1]}%
	\providecommand \BibitemOpen [0]{}%
	\providecommand \bibitemStop [0]{}%
	\providecommand \bibitemNoStop [0]{.\EOS\space}%
	\providecommand \EOS [0]{\spacefactor3000\relax}%
	\providecommand \BibitemShut  [1]{\csname bibitem#1\endcsname}%
	\let\auto@bib@innerbib\@empty
	\bibitem [{\citenamefont {Abe}\ \emph {et~al.}(2021)\citenamefont {Abe} \emph
		{et~al.}}]{PhysRevD.103.L011101}%
	\BibitemOpen
	\bibfield  {author} {\bibinfo {author} {\bibfnamefont {K.}~\bibnamefont
			{Abe}} \emph {et~al.} (\bibinfo {collaboration} {The T2K Collaboration}),\
	}\href {\doibase 10.1103/PhysRevD.103.L011101} {\bibfield  {journal}
		{\bibinfo  {journal} {Phys. Rev. D}\ }\textbf {\bibinfo {volume} {103}},\
		\bibinfo {pages} {L011101} (\bibinfo {year} {2021})}\BibitemShut {NoStop}%
	\bibitem [{\citenamefont {Acero}\ \emph {et~al.}(2019)\citenamefont {Acero}
		\emph {et~al.}}]{PhysRevLett.123.151803}%
	\BibitemOpen
	\bibfield  {author} {\bibinfo {author} {\bibfnamefont {M.~A.}\ \bibnamefont
			{Acero}} \emph {et~al.} (\bibinfo {collaboration} {NOvA Collaboration}),\
	}\href {\doibase 10.1103/PhysRevLett.123.151803} {\bibfield  {journal}
		{\bibinfo  {journal} {Phys. Rev. Lett.}\ }\textbf {\bibinfo {volume} {123}},\
		\bibinfo {pages} {151803} (\bibinfo {year} {2019})}\BibitemShut {NoStop}%
	\bibitem [{\citenamefont {Pontecorvo}()}]{osti_4349231}%
	\BibitemOpen
	\bibfield  {author} {\bibinfo {author} {\bibfnamefont {B.}~\bibnamefont
			{Pontecorvo}},\ }\href {https://www.osti.gov/biblio/4349231} {\bibinfo
		{journal} {Zhur. Eksptl'. i Teoret. Fiz.}\ }\BibitemShut {NoStop}%
	\bibitem [{\citenamefont {Maki}\ \emph {et~al.}(1962)\citenamefont {Maki} \emph
		{et~al.}}]{10.1143/PTP.28.870}%
	\BibitemOpen
	\bibfield  {journal} {  }\bibfield  {author} {\bibinfo {author} {\bibfnamefont
			{Z.}~\bibnamefont {Maki}} \emph {et~al.},\ }\href {\doibase
		10.1143/PTP.28.870} {\bibfield  {journal} {\bibinfo  {journal} {Progress of
				Theoretical Physics}\ }\textbf {\bibinfo {volume} {28}},\ \bibinfo {pages}
		{870} (\bibinfo {year} {1962})}\BibitemShut {NoStop}%
	\bibitem [{\citenamefont {Altarelli}\ and\ \citenamefont
		{Feruglio}(2010)}]{RevModPhys.82.2701}%
	\BibitemOpen
	\bibfield  {author} {\bibinfo {author} {\bibfnamefont {G.}~\bibnamefont
			{Altarelli}}\ and\ \bibinfo {author} {\bibfnamefont {F.}~\bibnamefont
			{Feruglio}},\ }\href {\doibase 10.1103/RevModPhys.82.2701} {\bibfield
		{journal} {\bibinfo  {journal} {Rev. Mod. Phys.}\ }\textbf {\bibinfo {volume}
			{82}},\ \bibinfo {pages} {2701} (\bibinfo {year} {2010})}\BibitemShut
	{NoStop}%
	\bibitem [{\citenamefont {Aartsen}\ \emph {et~al.}(2020)\citenamefont {Aartsen}
		\emph {et~al.}}]{PhysRevD.101.032006}%
	\BibitemOpen
	\bibfield  {author} {\bibinfo {author} {\bibfnamefont {M.~G.}\ \bibnamefont
			{Aartsen}} \emph {et~al.} (\bibinfo {collaboration} {IceCube-Gen2
			Collaboration and JUNO Collaboration Members}),\ }\href {\doibase
		10.1103/PhysRevD.101.032006} {\bibfield  {journal} {\bibinfo  {journal}
			{Phys. Rev. D}\ }\textbf {\bibinfo {volume} {101}},\ \bibinfo {pages}
		{032006} (\bibinfo {year} {2020})}\BibitemShut {NoStop}%
	\bibitem [{\citenamefont {Abe}\ \emph {et~al.}(2020)\citenamefont {Abe} \emph
		{et~al.}}]{Abe2020}%
	\BibitemOpen
	\bibfield  {author} {\bibinfo {author} {\bibfnamefont {K.}~\bibnamefont
			{Abe}} \emph {et~al.} (\bibinfo {collaboration} {The T2K Collaboration}),\
	}\href {\doibase 10.1038/s41586-020-2177-0} {\bibfield  {journal} {\bibinfo
			{journal} {Nature}\ }\textbf {\bibinfo {volume} {580}},\ \bibinfo {pages}
		{339} (\bibinfo {year} {2020})}\BibitemShut {NoStop}%
	\bibitem [{\citenamefont {Denton}\ \emph {et~al.}(2020)\citenamefont {Denton}
		\emph {et~al.}}]{PhysRevD.101.093001}%
	\BibitemOpen
	\bibfield  {author} {\bibinfo {author} {\bibfnamefont {P.~B.}\ \bibnamefont
			{Denton}} \emph {et~al.},\ }\href {\doibase 10.1103/PhysRevD.101.093001}
	{\bibfield  {journal} {\bibinfo  {journal} {Phys. Rev. D}\ }\textbf {\bibinfo
			{volume} {101}},\ \bibinfo {pages} {093001} (\bibinfo {year}
		{2020})}\BibitemShut {NoStop}%
	\bibitem [{\citenamefont {Nunokawa}\ \emph {et~al.}(2005)\citenamefont
		{Nunokawa} \emph {et~al.}}]{PhysRevD.72.013009}%
	\BibitemOpen
	\bibfield  {author} {\bibinfo {author} {\bibfnamefont {H.}~\bibnamefont
			{Nunokawa}} \emph {et~al.},\ }\href {\doibase 10.1103/PhysRevD.72.013009}
	{\bibfield  {journal} {\bibinfo  {journal} {Phys. Rev. D}\ }\textbf {\bibinfo
			{volume} {72}},\ \bibinfo {pages} {013009} (\bibinfo {year}
		{2005})}\BibitemShut {NoStop}%
	\bibitem [{\citenamefont {Qian}\ \emph {et~al.}(2012)\citenamefont {Qian} \emph
		{et~al.}}]{PhysRevD.86.113011}%
	\BibitemOpen
	\bibfield  {author} {\bibinfo {author} {\bibfnamefont {X.}~\bibnamefont
			{Qian}} \emph {et~al.},\ }\href {\doibase 10.1103/PhysRevD.86.113011}
	{\bibfield  {journal} {\bibinfo  {journal} {Phys. Rev. D}\ }\textbf {\bibinfo
			{volume} {86}},\ \bibinfo {pages} {113011} (\bibinfo {year}
		{2012})}\BibitemShut {NoStop}%
	\bibitem [{\citenamefont {Stanco}\ \emph {et~al.}(2017)\citenamefont {Stanco}
		\emph {et~al.}}]{PhysRevD.95.053002}%
	\BibitemOpen
	\bibfield  {author} {\bibinfo {author} {\bibfnamefont {L.}~\bibnamefont
			{Stanco}} \emph {et~al.},\ }\href {\doibase 10.1103/PhysRevD.95.053002}
	{\bibfield  {journal} {\bibinfo  {journal} {Phys. Rev. D}\ }\textbf {\bibinfo
			{volume} {95}},\ \bibinfo {pages} {053002} (\bibinfo {year}
		{2017})}\BibitemShut {NoStop}%
	\bibitem [{\citenamefont {de~Gouv\^ea}\ \emph {et~al.}(2021)\citenamefont
		{de~Gouv\^ea} \emph {et~al.}}]{PhysRevD.104.015038}%
	\BibitemOpen
	\bibfield  {author} {\bibinfo {author} {\bibfnamefont {A.}~\bibnamefont
			{de~Gouv\^ea}} \emph {et~al.},\ }\href {\doibase 10.1103/PhysRevD.104.015038}
	{\bibfield  {journal} {\bibinfo  {journal} {Phys. Rev. D}\ }\textbf {\bibinfo
			{volume} {104}},\ \bibinfo {pages} {015038} (\bibinfo {year}
		{2021})}\BibitemShut {NoStop}%
	\bibitem [{\citenamefont {Pancharatnam}(1956)}]{Pancharatnam1956}%
	\BibitemOpen
	\bibfield  {author} {\bibinfo {author} {\bibfnamefont {S.}~\bibnamefont
			{Pancharatnam}},\ }\href {\doibase 10.1007/BF03046050} {\bibfield  {journal}
		{\bibinfo  {journal} {Proceedings of the Indian Academy of Sciences - Section
				A}\ }\textbf {\bibinfo {volume} {44}},\ \bibinfo {pages} {247} (\bibinfo
		{year} {1956})}\BibitemShut {NoStop}%
	\bibitem [{\citenamefont {Berry}(1984)}]{doi:10.1098/rspa.1984.0023}%
	\BibitemOpen
	\bibfield  {author} {\bibinfo {author} {\bibfnamefont {M.~V.}\ \bibnamefont
			{Berry}},\ }\href {\doibase 10.1098/rspa.1984.0023} {\bibfield  {journal}
		{\bibinfo  {journal} {Proceedings of the Royal Society of London. A.
				Mathematical and Physical Sciences}\ }\textbf {\bibinfo {volume} {392}},\
		\bibinfo {pages} {45} (\bibinfo {year} {1984})}\BibitemShut {NoStop}%
	\bibitem [{\citenamefont {Ramaseshan}\ and\ \citenamefont
		{Nityananda}(1986)}]{10.2307/24090242}%
	\BibitemOpen
	\bibfield  {author} {\bibinfo {author} {\bibfnamefont {S.}~\bibnamefont
			{Ramaseshan}}\ and\ \bibinfo {author} {\bibfnamefont {R.}~\bibnamefont
			{Nityananda}},\ }\href {http://www.jstor.org/stable/24090242} {\bibfield
		{journal} {\bibinfo  {journal} {Current Science}\ }\textbf {\bibinfo {volume}
			{55}},\ \bibinfo {pages} {1225} (\bibinfo {year} {1986})}\BibitemShut
	{NoStop}%
	\bibitem [{\citenamefont {Berry}(1987)}]{doi:10.1080/09500348714551321}%
	\BibitemOpen
	\bibfield  {author} {\bibinfo {author} {\bibfnamefont {M.}~\bibnamefont
			{Berry}},\ }\href {\doibase 10.1080/09500348714551321} {\bibfield  {journal}
		{\bibinfo  {journal} {Journal of Modern Optics}\ }\textbf {\bibinfo {volume}
			{34}},\ \bibinfo {pages} {1401} (\bibinfo {year} {1987})}\BibitemShut
	{NoStop}%
	\bibitem [{\citenamefont {Aharonov}\ and\ \citenamefont
		{Anandan}(1987)}]{PhysRevLett.58.1593}%
	\BibitemOpen
	\bibfield  {author} {\bibinfo {author} {\bibfnamefont {Y.}~\bibnamefont
			{Aharonov}}\ and\ \bibinfo {author} {\bibfnamefont {J.}~\bibnamefont
			{Anandan}},\ }\href {\doibase 10.1103/PhysRevLett.58.1593} {\bibfield
		{journal} {\bibinfo  {journal} {Phys. Rev. Lett.}\ }\textbf {\bibinfo
			{volume} {58}},\ \bibinfo {pages} {1593} (\bibinfo {year}
		{1987})}\BibitemShut {NoStop}%
	\bibitem [{\citenamefont {Samuel}\ and\ \citenamefont
		{Bhandari}(1988)}]{PhysRevLett.60.2339}%
	\BibitemOpen
	\bibfield  {author} {\bibinfo {author} {\bibfnamefont {J.}~\bibnamefont
			{Samuel}}\ and\ \bibinfo {author} {\bibfnamefont {R.}~\bibnamefont
			{Bhandari}},\ }\href {\doibase 10.1103/PhysRevLett.60.2339} {\bibfield
		{journal} {\bibinfo  {journal} {Phys. Rev. Lett.}\ }\textbf {\bibinfo
			{volume} {60}},\ \bibinfo {pages} {2339} (\bibinfo {year}
		{1988})}\BibitemShut {NoStop}%
	\bibitem [{\citenamefont {Cohen}\ \emph {et~al.}(2019)\citenamefont {Cohen}
		\emph {et~al.}}]{Cohen2019}%
	\BibitemOpen
	\bibfield  {author} {\bibinfo {author} {\bibfnamefont {E.}~\bibnamefont
			{Cohen}} \emph {et~al.},\ }\href {\doibase 10.1038/s42254-019-0071-1}
	{\bibfield  {journal} {\bibinfo  {journal} {Nature Reviews Physics}\ }\textbf
		{\bibinfo {volume} {1}},\ \bibinfo {pages} {437} (\bibinfo {year}
		{2019})}\BibitemShut {NoStop}%
	\bibitem [{\citenamefont {Jisha}\ \emph {et~al.}()\citenamefont {Jisha} \emph
		{et~al.}}]{https://doi.org/10.1002/lpor.202100003}%
	\BibitemOpen
	\bibfield  {author} {\bibinfo {author} {\bibfnamefont {C.~P.}\ \bibnamefont
			{Jisha}} \emph {et~al.},\ }\href {\doibase
		https://doi.org/10.1002/lpor.202100003} {\bibinfo  {journal} {Laser \&
			Photonics Reviews}\ ,\ \bibinfo {pages} {2100003}}\BibitemShut {NoStop}%
	\bibitem [{\citenamefont {Nakagawa}(1987)}]{NAKAGAWA1987145}%
	\BibitemOpen
	\bibfield  {journal} {  }\bibfield  {author} {\bibinfo {author} {\bibfnamefont
			{N.}~\bibnamefont {Nakagawa}},\ }\href {\doibase
		https://doi.org/10.1016/S0003-4916(87)80007-6} {\bibfield  {journal}
		{\bibinfo  {journal} {Annals of Physics}\ }\textbf {\bibinfo {volume}
			{179}},\ \bibinfo {pages} {145} (\bibinfo {year} {1987})}\BibitemShut
	{NoStop}%
	\bibitem [{\citenamefont {Naumov}(1994)}]{NAUMOV1994351}%
	\BibitemOpen
	\bibfield  {author} {\bibinfo {author} {\bibfnamefont {V.~A.}\ \bibnamefont
			{Naumov}},\ }\href {\doibase https://doi.org/10.1016/0370-2693(94)91231-9}
	{\bibfield  {journal} {\bibinfo  {journal} {Physics Letters B}\ }\textbf
		{\bibinfo {volume} {323}},\ \bibinfo {pages} {351} (\bibinfo {year}
		{1994})}\BibitemShut {NoStop}%
	\bibitem [{\citenamefont {He}\ \emph {et~al.}(2005)\citenamefont {He} \emph
		{et~al.}}]{PhysRevD.72.053012}%
	\BibitemOpen
	\bibfield  {author} {\bibinfo {author} {\bibfnamefont {X.-G.}\ \bibnamefont
			{He}} \emph {et~al.},\ }\href {\doibase 10.1103/PhysRevD.72.053012}
	{\bibfield  {journal} {\bibinfo  {journal} {Phys. Rev. D}\ }\textbf {\bibinfo
			{volume} {72}},\ \bibinfo {pages} {053012} (\bibinfo {year}
		{2005})}\BibitemShut {NoStop}%
	\bibitem [{\citenamefont {Joshi}\ and\ \citenamefont
		{Jain}(2016)}]{JOSHI2016135}%
	\BibitemOpen
	\bibfield  {author} {\bibinfo {author} {\bibfnamefont {S.}~\bibnamefont
			{Joshi}}\ and\ \bibinfo {author} {\bibfnamefont {S.~R.}\ \bibnamefont
			{Jain}},\ }\href {\doibase https://doi.org/10.1016/j.physletb.2016.01.023}
	{\bibfield  {journal} {\bibinfo  {journal} {Physics Letters B}\ }\textbf
		{\bibinfo {volume} {754}},\ \bibinfo {pages} {135} (\bibinfo {year}
		{2016})}\BibitemShut {NoStop}%
	\bibitem [{\citenamefont {Johns}\ and\ \citenamefont
		{Fuller}(2017)}]{PhysRevD.95.043003}%
	\BibitemOpen
	\bibfield  {author} {\bibinfo {author} {\bibfnamefont {L.}~\bibnamefont
			{Johns}}\ and\ \bibinfo {author} {\bibfnamefont {G.~M.}\ \bibnamefont
			{Fuller}},\ }\href {\doibase 10.1103/PhysRevD.95.043003} {\bibfield
		{journal} {\bibinfo  {journal} {Phys. Rev. D}\ }\textbf {\bibinfo {volume}
			{95}},\ \bibinfo {pages} {043003} (\bibinfo {year} {2017})}\BibitemShut
	{NoStop}%
	\bibitem [{\citenamefont {Joshi}\ and\ \citenamefont
		{Jain}(2017)}]{PhysRevD.96.096004}%
	\BibitemOpen
	\bibfield  {author} {\bibinfo {author} {\bibfnamefont {S.}~\bibnamefont
			{Joshi}}\ and\ \bibinfo {author} {\bibfnamefont {S.~R.}\ \bibnamefont
			{Jain}},\ }\href {\doibase 10.1103/PhysRevD.96.096004} {\bibfield  {journal}
		{\bibinfo  {journal} {Phys. Rev. D}\ }\textbf {\bibinfo {volume} {96}},\
		\bibinfo {pages} {096004} (\bibinfo {year} {2017})}\BibitemShut {NoStop}%
	\bibitem [{\citenamefont {Simon}\ and\ \citenamefont
		{Mukunda}(1993)}]{PhysRevLett.70.880}%
	\BibitemOpen
	\bibfield  {author} {\bibinfo {author} {\bibfnamefont {R.}~\bibnamefont
			{Simon}}\ and\ \bibinfo {author} {\bibfnamefont {N.}~\bibnamefont
			{Mukunda}},\ }\href {\doibase 10.1103/PhysRevLett.70.880} {\bibfield
		{journal} {\bibinfo  {journal} {Phys. Rev. Lett.}\ }\textbf {\bibinfo
			{volume} {70}},\ \bibinfo {pages} {880} (\bibinfo {year} {1993})}\BibitemShut
	{NoStop}%
	\bibitem [{\citenamefont {Mukunda}\ and\ \citenamefont
		{Simon}(1993)}]{MUKUNDA1993205}%
	\BibitemOpen
	\bibfield  {author} {\bibinfo {author} {\bibfnamefont {N.}~\bibnamefont
			{Mukunda}}\ and\ \bibinfo {author} {\bibfnamefont {R.}~\bibnamefont
			{Simon}},\ }\href {\doibase https://doi.org/10.1006/aphy.1993.1093}
	{\bibfield  {journal} {\bibinfo  {journal} {Annals of Physics}\ }\textbf
		{\bibinfo {volume} {228}},\ \bibinfo {pages} {205} (\bibinfo {year}
		{1993})}\BibitemShut {NoStop}%
	\bibitem [{\citenamefont {Rabei}\ \emph {et~al.}(1999)\citenamefont {Rabei}
		\emph {et~al.}}]{PhysRevA.60.3397}%
	\BibitemOpen
	\bibfield  {author} {\bibinfo {author} {\bibfnamefont {E.~M.}\ \bibnamefont
			{Rabei}} \emph {et~al.},\ }\href {\doibase 10.1103/PhysRevA.60.3397}
	{\bibfield  {journal} {\bibinfo  {journal} {Phys. Rev. A}\ }\textbf {\bibinfo
			{volume} {60}},\ \bibinfo {pages} {3397} (\bibinfo {year}
		{1999})}\BibitemShut {NoStop}%
	\bibitem [{\citenamefont {Blasone}\ \emph {et~al.}(1999)\citenamefont {Blasone}
		\emph {et~al.}}]{BLASONE1999262}%
	\BibitemOpen
	\bibfield  {author} {\bibinfo {author} {\bibfnamefont {M.}~\bibnamefont
			{Blasone}} \emph {et~al.},\ }\href {\doibase
		https://doi.org/10.1016/S0370-2693(99)01137-5} {\bibfield  {journal}
		{\bibinfo  {journal} {Physics Letters B}\ }\textbf {\bibinfo {volume}
			{466}},\ \bibinfo {pages} {262} (\bibinfo {year} {1999})}\BibitemShut
	{NoStop}%
	\bibitem [{\citenamefont {Aharonov}\ and\ \citenamefont
		{Bohm}(1959)}]{PhysRev.115.485}%
	\BibitemOpen
	\bibfield  {author} {\bibinfo {author} {\bibfnamefont {Y.}~\bibnamefont
			{Aharonov}}\ and\ \bibinfo {author} {\bibfnamefont {D.}~\bibnamefont
			{Bohm}},\ }\href {\doibase 10.1103/PhysRev.115.485} {\bibfield  {journal}
		{\bibinfo  {journal} {Phys. Rev.}\ }\textbf {\bibinfo {volume} {115}},\
		\bibinfo {pages} {485} (\bibinfo {year} {1959})}\BibitemShut {NoStop}%
	\bibitem [{\citenamefont {Wang}\ \emph {et~al.}(2001)\citenamefont {Wang} \emph
		{et~al.}}]{PhysRevD.63.053003}%
	\BibitemOpen
	\bibfield  {author} {\bibinfo {author} {\bibfnamefont {X.-B.}\ \bibnamefont
			{Wang}} \emph {et~al.},\ }\href {\doibase 10.1103/PhysRevD.63.053003}
	{\bibfield  {journal} {\bibinfo  {journal} {Phys. Rev. D}\ }\textbf {\bibinfo
			{volume} {63}},\ \bibinfo {pages} {053003} (\bibinfo {year}
		{2001})}\BibitemShut {NoStop}%
	\bibitem [{\citenamefont {Dixit}\ \emph {et~al.}(2018)\citenamefont {Dixit},
		\emph {et~al.}}]{Dixit_2018}%
	\BibitemOpen
	\bibfield  {author} {\bibinfo {author} {\bibfnamefont {K.}~\bibnamefont
			{Dixit}}, ,  \emph {et~al.},\ }\href {\doibase 10.1088/1361-6471/aac454}
	{\bibfield  {journal} {\bibinfo  {journal} {Journal of Physics G: Nuclear and
				Particle Physics}\ }\textbf {\bibinfo {volume} {45}},\ \bibinfo {pages}
		{085002} (\bibinfo {year} {2018})}\BibitemShut {NoStop}%
	\bibitem [{\citenamefont {Mehta}(2009)}]{PhysRevD.79.096013}%
	\BibitemOpen
	\bibfield  {author} {\bibinfo {author} {\bibfnamefont {P.}~\bibnamefont
			{Mehta}},\ }\href {\doibase 10.1103/PhysRevD.79.096013} {\bibfield  {journal}
		{\bibinfo  {journal} {Phys. Rev. D}\ }\textbf {\bibinfo {volume} {79}},\
		\bibinfo {pages} {096013} (\bibinfo {year} {2009})}\BibitemShut {NoStop}%
	\bibitem [{\citenamefont {Fuentes-Guridi}\ \emph {et~al.}(2002)\citenamefont
		{Fuentes-Guridi} \emph {et~al.}}]{PhysRevLett.89.220404}%
	\BibitemOpen
	\bibfield  {author} {\bibinfo {author} {\bibfnamefont {I.}~\bibnamefont
			{Fuentes-Guridi}} \emph {et~al.},\ }\href {\doibase
		10.1103/PhysRevLett.89.220404} {\bibfield  {journal} {\bibinfo  {journal}
			{Phys. Rev. Lett.}\ }\textbf {\bibinfo {volume} {89}},\ \bibinfo {pages}
		{220404} (\bibinfo {year} {2002})}\BibitemShut {NoStop}%
	\bibitem [{\citenamefont {Larson}(2012)}]{PhysRevLett.108.033601}%
	\BibitemOpen
	\bibfield  {author} {\bibinfo {author} {\bibfnamefont {J.}~\bibnamefont
			{Larson}},\ }\href {\doibase 10.1103/PhysRevLett.108.033601} {\bibfield
		{journal} {\bibinfo  {journal} {Phys. Rev. Lett.}\ }\textbf {\bibinfo
			{volume} {108}},\ \bibinfo {pages} {033601} (\bibinfo {year}
		{2012})}\BibitemShut {NoStop}%
	\bibitem [{\citenamefont {Calder\'on}\ and\ \citenamefont
		{De~Zela}(2016)}]{PhysRevA.93.033823}%
	\BibitemOpen
	\bibfield  {author} {\bibinfo {author} {\bibfnamefont {J.}~\bibnamefont
			{Calder\'on}}\ and\ \bibinfo {author} {\bibfnamefont {F.}~\bibnamefont
			{De~Zela}},\ }\href {\doibase 10.1103/PhysRevA.93.033823} {\bibfield
		{journal} {\bibinfo  {journal} {Phys. Rev. A}\ }\textbf {\bibinfo {volume}
			{93}},\ \bibinfo {pages} {033823} (\bibinfo {year} {2016})}\BibitemShut
	{NoStop}%
	\bibitem [{\citenamefont {Gasparinetti}\ \emph {et~al.}(2016)\citenamefont
		{Gasparinetti} \emph {et~al.}}]{Gasparinettie1501732}%
	\BibitemOpen
	\bibfield  {author} {\bibinfo {author} {\bibfnamefont {S.}~\bibnamefont
			{Gasparinetti}} \emph {et~al.},\ }\href {\doibase 10.1126/sciadv.1501732}
	{\bibfield  {journal} {\bibinfo  {journal} {Science Advances}\ }\textbf
		{\bibinfo {volume} {2}} (\bibinfo {year} {2016}),\
		10.1126/sciadv.1501732}\BibitemShut {NoStop}%
	\bibitem [{\citenamefont {Uhlmann}(1986)}]{UHLMANN1986229}%
	\BibitemOpen
	\bibfield  {author} {\bibinfo {author} {\bibfnamefont {A.}~\bibnamefont
			{Uhlmann}},\ }\href {\doibase https://doi.org/10.1016/0034-4877(86)90055-8}
	{\bibfield  {journal} {\bibinfo  {journal} {Reports on Mathematical Physics}\
		}\textbf {\bibinfo {volume} {24}},\ \bibinfo {pages} {229} (\bibinfo {year}
		{1986})}\BibitemShut {NoStop}%
	\bibitem [{\citenamefont {Sj\"oqvist}\ \emph {et~al.}(2000)\citenamefont
		{Sj\"oqvist} \emph {et~al.}}]{PhysRevLett.85.2845}%
	\BibitemOpen
	\bibfield  {author} {\bibinfo {author} {\bibfnamefont {E.}~\bibnamefont
			{Sj\"oqvist}} \emph {et~al.},\ }\href {\doibase 10.1103/PhysRevLett.85.2845}
	{\bibfield  {journal} {\bibinfo  {journal} {Phys. Rev. Lett.}\ }\textbf
		{\bibinfo {volume} {85}},\ \bibinfo {pages} {2845} (\bibinfo {year}
		{2000})}\BibitemShut {NoStop}%
	\bibitem [{\citenamefont {Carollo}\ \emph {et~al.}(2003)\citenamefont {Carollo}
		\emph {et~al.}}]{PhysRevLett.90.160402}%
	\BibitemOpen
	\bibfield  {author} {\bibinfo {author} {\bibfnamefont {A.}~\bibnamefont
			{Carollo}} \emph {et~al.},\ }\href {\doibase 10.1103/PhysRevLett.90.160402}
	{\bibfield  {journal} {\bibinfo  {journal} {Phys. Rev. Lett.}\ }\textbf
		{\bibinfo {volume} {90}},\ \bibinfo {pages} {160402} (\bibinfo {year}
		{2003})}\BibitemShut {NoStop}%
	\bibitem [{\citenamefont {Sj\"oqvist}(2000)}]{PhysRevA.62.022109}%
	\BibitemOpen
	\bibfield  {author} {\bibinfo {author} {\bibfnamefont {E.}~\bibnamefont
			{Sj\"oqvist}},\ }\href {\doibase 10.1103/PhysRevA.62.022109} {\bibfield
		{journal} {\bibinfo  {journal} {Phys. Rev. A}\ }\textbf {\bibinfo {volume}
			{62}},\ \bibinfo {pages} {022109} (\bibinfo {year} {2000})}\BibitemShut
	{NoStop}%
	\bibitem [{\citenamefont {Yi}\ \emph {et~al.}(2004)\citenamefont {Yi} \emph
		{et~al.}}]{PhysRevLett.92.150406}%
	\BibitemOpen
	\bibfield  {author} {\bibinfo {author} {\bibfnamefont {X.~X.}\ \bibnamefont
			{Yi}} \emph {et~al.},\ }\href {\doibase 10.1103/PhysRevLett.92.150406}
	{\bibfield  {journal} {\bibinfo  {journal} {Phys. Rev. Lett.}\ }\textbf
		{\bibinfo {volume} {92}},\ \bibinfo {pages} {150406} (\bibinfo {year}
		{2004})}\BibitemShut {NoStop}%
	\bibitem [{\citenamefont {Leek}\ \emph {et~al.}(2007)\citenamefont {Leek} \emph
		{et~al.}}]{Leek1889}%
	\BibitemOpen
	\bibfield  {author} {\bibinfo {author} {\bibfnamefont {P.~J.}\ \bibnamefont
			{Leek}} \emph {et~al.},\ }\href {\doibase 10.1126/science.1149858} {\bibfield
		{journal} {\bibinfo  {journal} {Science}\ }\textbf {\bibinfo {volume}
			{318}},\ \bibinfo {pages} {1889} (\bibinfo {year} {2007})}\BibitemShut
	{NoStop}%
	\bibitem [{\citenamefont {Tong}\ \emph {et~al.}(2004)\citenamefont {Tong} \emph
		{et~al.}}]{PhysRevLett.93.080405}%
	\BibitemOpen
	\bibfield  {author} {\bibinfo {author} {\bibfnamefont {D.~M.}\ \bibnamefont
			{Tong}} \emph {et~al.},\ }\href {\doibase 10.1103/PhysRevLett.93.080405}
	{\bibfield  {journal} {\bibinfo  {journal} {Phys. Rev. Lett.}\ }\textbf
		{\bibinfo {volume} {93}},\ \bibinfo {pages} {080405} (\bibinfo {year}
		{2004})}\BibitemShut {NoStop}%
	\bibitem [{\citenamefont {Mukunda}\ \emph {et~al.}(2001)\citenamefont {Mukunda}
		\emph {et~al.}}]{PhysRevA.65.012102}%
	\BibitemOpen
	\bibfield  {author} {\bibinfo {author} {\bibfnamefont {N.}~\bibnamefont
			{Mukunda}} \emph {et~al.},\ }\href {\doibase 10.1103/PhysRevA.65.012102}
	{\bibfield  {journal} {\bibinfo  {journal} {Phys. Rev. A}\ }\textbf {\bibinfo
			{volume} {65}},\ \bibinfo {pages} {012102} (\bibinfo {year}
		{2001})}\BibitemShut {NoStop}%
	\bibitem [{\citenamefont {Khan}\ \emph {et~al.}(2021)\citenamefont {Khan} \emph
		{et~al.}}]{KHAN2021100061}%
	\BibitemOpen
	\bibfield  {author} {\bibinfo {author} {\bibfnamefont {M.~N.}\ \bibnamefont
			{Khan}} \emph {et~al.},\ }\href {\doibase
		https://doi.org/10.1016/j.physo.2021.100061} {\bibfield  {journal} {\bibinfo
			{journal} {Physics Open}\ }\textbf {\bibinfo {volume} {7}},\ \bibinfo {pages}
		{100061} (\bibinfo {year} {2021})}\BibitemShut {NoStop}%
	\bibitem [{\citenamefont {Manini}\ and\ \citenamefont
		{Pistolesi}(2000)}]{PhysRevLett.85.3067}%
	\BibitemOpen
	\bibfield  {author} {\bibinfo {author} {\bibfnamefont {N.}~\bibnamefont
			{Manini}}\ and\ \bibinfo {author} {\bibfnamefont {F.}~\bibnamefont
			{Pistolesi}},\ }\href {\doibase 10.1103/PhysRevLett.85.3067} {\bibfield
		{journal} {\bibinfo  {journal} {Phys. Rev. Lett.}\ }\textbf {\bibinfo
			{volume} {85}},\ \bibinfo {pages} {3067} (\bibinfo {year}
		{2000})}\BibitemShut {NoStop}%
	\bibitem [{\citenamefont {Filipp}\ and\ \citenamefont
		{Sj\"oqvist}(2003)}]{PhysRevLett.90.050403}%
	\BibitemOpen
	\bibfield  {author} {\bibinfo {author} {\bibfnamefont {S.}~\bibnamefont
			{Filipp}}\ and\ \bibinfo {author} {\bibfnamefont {E.}~\bibnamefont
			{Sj\"oqvist}},\ }\href {\doibase 10.1103/PhysRevLett.90.050403} {\bibfield
		{journal} {\bibinfo  {journal} {Phys. Rev. Lett.}\ }\textbf {\bibinfo
			{volume} {90}},\ \bibinfo {pages} {050403} (\bibinfo {year}
		{2003})}\BibitemShut {NoStop}%
	\bibitem [{\citenamefont {Wong}\ \emph {et~al.}(2005)\citenamefont {Wong} \emph
		{et~al.}}]{PhysRevLett.94.070406}%
	\BibitemOpen
	\bibfield  {author} {\bibinfo {author} {\bibfnamefont {H.~M.}\ \bibnamefont
			{Wong}} \emph {et~al.},\ }\href {\doibase 10.1103/PhysRevLett.94.070406}
	{\bibfield  {journal} {\bibinfo  {journal} {Phys. Rev. Lett.}\ }\textbf
		{\bibinfo {volume} {94}},\ \bibinfo {pages} {070406} (\bibinfo {year}
		{2005})}\BibitemShut {NoStop}%
	\bibitem [{\citenamefont {Kobayashi}\ and\ \citenamefont
		{Maskawa}(1973)}]{10.1143/PTP.49.652}%
	\BibitemOpen
	\bibfield  {author} {\bibinfo {author} {\bibfnamefont {M.}~\bibnamefont
			{Kobayashi}}\ and\ \bibinfo {author} {\bibfnamefont {T.}~\bibnamefont
			{Maskawa}},\ }\href {\doibase 10.1143/PTP.49.652} {\bibfield  {journal}
		{\bibinfo  {journal} {Progress of Theoretical Physics}\ }\textbf {\bibinfo
			{volume} {49}},\ \bibinfo {pages} {652} (\bibinfo {year} {1973})}\BibitemShut
	{NoStop}%
	\bibitem [{\citenamefont {Wolfenstein}(1978)}]{PhysRevD.17.2369}%
	\BibitemOpen
	\bibfield  {author} {\bibinfo {author} {\bibfnamefont {L.}~\bibnamefont
			{Wolfenstein}},\ }\href {\doibase 10.1103/PhysRevD.17.2369} {\bibfield
		{journal} {\bibinfo  {journal} {Phys. Rev. D}\ }\textbf {\bibinfo {volume}
			{17}},\ \bibinfo {pages} {2369} (\bibinfo {year} {1978})}\BibitemShut
	{NoStop}%
	\bibitem [{\citenamefont {Mikheyev}\ and\ \citenamefont
		{Smirnov}(1986)}]{Mikheyev1986}%
	\BibitemOpen
	\bibfield  {author} {\bibinfo {author} {\bibfnamefont {S.~P.}\ \bibnamefont
			{Mikheyev}}\ and\ \bibinfo {author} {\bibfnamefont {A.~Y.}\ \bibnamefont
			{Smirnov}},\ }\href {\doibase 10.1007/BF02508049} {\bibfield  {journal}
		{\bibinfo  {journal} {Il Nuovo Cimento C}\ }\textbf {\bibinfo {volume} {9}},\
		\bibinfo {pages} {17} (\bibinfo {year} {1986})}\BibitemShut {NoStop}%
	\bibitem [{\citenamefont {Capolupo}\ \emph {et~al.}(2018)\citenamefont
		{Capolupo} \emph {et~al.}}]{CAPOLUPO2018216}%
	\BibitemOpen
	\bibfield  {author} {\bibinfo {author} {\bibfnamefont {A.}~\bibnamefont
			{Capolupo}} \emph {et~al.},\ }\href {\doibase
		https://doi.org/10.1016/j.physletb.2018.03.016} {\bibfield  {journal}
		{\bibinfo  {journal} {Physics Letters B}\ }\textbf {\bibinfo {volume}
			{780}},\ \bibinfo {pages} {216} (\bibinfo {year} {2018})}\BibitemShut
	{NoStop}%
	\bibitem [{\citenamefont {Lu}(2021)}]{LU2021136376}%
	\BibitemOpen
	\bibfield  {author} {\bibinfo {author} {\bibfnamefont {J.}~\bibnamefont
			{Lu}},\ }\href {\doibase https://doi.org/10.1016/j.physletb.2021.136376}
	{\bibfield  {journal} {\bibinfo  {journal} {Physics Letters B}\ }\textbf
		{\bibinfo {volume} {818}},\ \bibinfo {pages} {136376} (\bibinfo {year}
		{2021})}\BibitemShut {NoStop}%
	\bibitem [{\citenamefont {Capolupo}\ \emph {et~al.}(2021)\citenamefont
		{Capolupo} \emph {et~al.}}]{Capolupo:2021enm}%
	\BibitemOpen
	\bibfield  {author} {\bibinfo {author} {\bibfnamefont {A.}~\bibnamefont
			{Capolupo}} \emph {et~al.},\ }\href@noop {} {\  (\bibinfo {year} {2021})},\
	\Eprint {http://arxiv.org/abs/2107.08719} {arXiv:2107.08719 [hep-ph]}
	\BibitemShut {NoStop}%
	\bibitem [{\citenamefont {Johns}(2021)}]{Johns:2021ets}%
	\BibitemOpen
	\bibfield  {author} {\bibinfo {author} {\bibfnamefont {L.}~\bibnamefont
			{Johns}},\ }\href@noop {} {\  (\bibinfo {year} {2021})},\ \Eprint
	{http://arxiv.org/abs/2107.11434} {arXiv:2107.11434 [hep-ph]} \BibitemShut
	{NoStop}%
	\bibitem [{\citenamefont {Zyla}\ \emph {et~al.}(2020)\citenamefont {Zyla} \emph
		{et~al.}}]{10.1093/ptep/ptaa104}%
	\BibitemOpen
	\bibfield  {author} {\bibinfo {author} {\bibfnamefont {P.~A.}\ \bibnamefont
			{Zyla}} \emph {et~al.} (\bibinfo {collaboration} {Particle Data Group}),\
	}\href {https://doi.org/10.1093/ptep/ptaa104} {\bibfield  {journal} {\bibinfo
			{journal} {Progress of Theoretical and Experimental Physics}\ }\textbf
		{\bibinfo {volume} {2020}} (\bibinfo {year} {2020})},\ \bibinfo {note}
	{083C01}\BibitemShut {NoStop}%
\end{thebibliography}
\end{document}